\documentclass[pre,twocolumn,nofootinbib,twoside,showpacs,superscriptaddress,tightenlines,floatfix,aps]{revtex4-2}
\usepackage{dcolumn}
\usepackage{lipsum}
\usepackage{graphicx,amssymb,amsmath,epsf,bm}
\usepackage[utf8x]{inputenc}
\usepackage[T1]{fontenc}
\usepackage[english]{babel} 
\usepackage{tikz} 
\usepackage[compat=1.1.0]{tikz-feynman}
\usepackage[export]{adjustbox} 
\usepackage[normalem]{ulem}
\usepackage{color}
\usepackage{mathtools}
\usepackage{dsfont}
\usepackage{nicefrac}
\usepackage{hyperref}
\hypersetup{
    allcolors=blue,
    colorlinks=true,
    bookmarks=true,
    pdfpagemode=FullScreen,
}

\usepackage{ulem}
\usepackage{tikz}
\usepackage{bm}
\usepackage{physics}
\usepackage{txfonts}
\usepackage{mathrsfs}
\usepackage{amsmath}
\usepackage{xcolor}
\usepackage{upgreek}

\definecolor{nred}{RGB}{224,0,0}
\definecolor{nblue}  {RGB}{28,130,185}
\definecolor{dgreen} {RGB}{38,238,21}
\definecolor{norange}{RGB}{230,120,20}

\begin{document}

\title{Critical Probability Distributions of the order parameter at two loops I: Ising universality class  }

\author{Sankarshan Sahu }
\affiliation{Sorbonne Universit\'e, CNRS, Laboratoire de Physique
Th\'eorique de la Mati\`ere Condens\'ee, LPTMC, 75005 Paris, France}
\date{\today}
\begin{abstract}
There exists an entire family of universal PDFs of the magnetization mode of the three dimensional Ising model parameterized by $\zeta = \lim_{L,\xi_{\infty}}L/\xi_{\infty}$ which is the ratio of the system size $L$ to the bulk correlation length $\xi_{\infty}$
with both the thermodynamic limit and the critical limit being taken simultaneously at fixed $\zeta$. Recently, the probability distribution functions (PDFs) of the magnetization mode of the three-dimensional Ising model has been computed at one-loop in the $\epsilon=4-d$ expansion [\textbf{arXiv preprint arXiv:2407.12603 (2024)}]. We show how these PDFs or, equivalently, the rate functions which are their logarithm, can be systematically computed at second order of the perturbative expansion. We compute the whole family of universal rate-functions and show that their agreement with the Monte Carlo data improves significantly at this order when compared to their one-loop counterpart.
\end{abstract}


\maketitle

\section{Introduction}
The speculation of the relationship between Renormalization Group and Probability Theory goes back at least as far as  1973 where Bleher and Sinai \cite{bleher1973investigation} hinted towards it in the context of hierarchical models which are simplified versions of the Ising and O$(n)$ models. This idea was further extended by Jona-Lasinio in  \cite{lona1975renormalization}, where he further established the connection between limiting theorems in probability theory and renormalization group. An example of such a limiting theorem is the Central Limit Theorem (CLT).

The Central Limit Theorem (CLT) is of crucial importance in many key areas of research in physics, mathematics, and quantitative finance. According to this theorem, for a large number $N$ of identically distributed independent random variables $\hat{\sigma}_{i}$, their sum $S=\sum_{i}\hat{\sigma}_{i}$ has fluctuations of order $\sqrt{N}$ and the asymptotic probability distribution of the properly normalized variable $S/\sqrt{N}$ is a Gaussian law with finite variance if both the mean and the variance of the $\hat{\sigma}_i$'s are finite. Most importantly, the resulting distribution (here, the Gaussian law) is universal, that is, independent of the probability distribution from which the variables $\hat{\sigma}_{i}$'s are chosen [except for the mean and variance]. The CLT itself has been generalized to the cases where the variables $\hat{\sigma}_{i}$ are weakly correlated (instead of being completely independent) \cite{botet2002universal, dedecker2007weak} . Physically, the picture becomes that of finite clusters of $\hat{\sigma}_{i}$ of size equal to their correlation length (say $\xi$) which are strongly correlated within the cluster, but these clusters again being independent of each other. However, no such limiting theorem exists in case of strongly correlated systems, that is, for systems where all the $\hat{\sigma}_{i}$'s are algebraically correlated.\\

 In our previous article \cite{sahu2024generalization}, we formulated a systematic perturbative approach to these limiting distributions for strongly correlated systems using the $\epsilon=4-d$ expansion and computed these PDFs at one-loop order for the $O(n)$ and Ising models. A calculation of these PDFs already exists in  a non-perturbative setting or more precisely at the leading order of the Functional Renormalization Group FRG calculations (LPA approximation) \cite{balog2022critical}. The goal of the current article is to go to the next order of the $\epsilon=4-d$ expansion, thereby performing a two-loop calculation of the PDFs. We do this once again in the context of the well known Ising model in three dimensions. The generalization to the $O(n)$ case is not as trivial as in the one-loop calculation  and is beyond the scope of the current work.

Before going further we must clarify what is to be gained from studying these probability distributions beyond mere academic curiosity. One of the greatest successes of Renormalization Group was the computation of universal critical exponents and critical amplitude ratios to striking accuracy in both the perturbative as well as non-perturbative setting. Most of the time, therefore, the characterization of the different universality classes is carried out using the values of the critical exponents, that is, by two numbers since only two of them are independent, at least as far as the leading scaling behavior is concerned.
However, for comparison with numerical simulations and experiments, it is much better to have universal functions rather than a small amount of numbers. There are two kinds of universal functions close to criticality, those that depend on wavenumbers, this is the case of the universal critical structure factor, and those that are field dependent such as universal finite-size scaling functions routinely computed in numerical simulations. We compute below one such function, namely the critical probability distribution function. Notice that both the computation and the measure of the critical structure factors are delicate whereas the critical PDFs are not as shown below. One must keep in mind that although universal, these critical PDFs have a dependence on boundary conditions \footnote{Here only regular isotropic lattices have been considered, or more precisely all isotropic lattices with same boundary conditions will yield the same PDF.} . In this paper, we only consider periodic boundary conditions.\\

 In our previous work \cite{sahu2024generalization} and in \cite{balog2022critical, antal2004probability}, it has already been shown that taking the critical limit, $T\to T_c$, and the thermodynamic limit simultaneously, one does not end up with just one PDF but a family of PDFs  indexed by $\zeta=L/\xi_{\infty}$, with $L$ being the system size and $\xi_{\infty}$ being the correlation length of the system depending on the way, the critical limit, and the thermodynamic limit is approached. Although the one-loop results are qualitatively good, they are very far from being quantitatively accurate. In the current article we show how going to two loops actually brings us to the level of quantitative accuracy when compared with the Monte-Carlo simulations, thereby tremendously improving the one-loop results.

\section{Field Theoretic Formalism}

 In the following, we consider the Ising model described by the Hamiltonian: 
 \begin{equation}
     \mathcal{H}=-J\sum_{<ij>}\hat{\sigma}_{i} \hat{\sigma}_{j},
 \end{equation}
 with coupling constant $J>0$, and spins $\hat{\sigma}_{i}=\pm 1$ where $\langle ij\rangle$ represent the nearest-neighbor interactions on a hyper cubic $d$ dimensional lattice with length $L$ and periodic boundary conditions. When $L\to\infty$, a second order phase transition occurs in the Ising model for some critical temperature $T_{c}$ and is characterized by the divergence of the correlation length $\xi_\infty$, i.e. at a temperature $T=T_{c}$, $\xi_{\infty}\rightarrow\infty$, which is the hallmark of strong correlations. 

What we are actually interested in is the PDF  of the total normalized spin defined as $\hat{s}=L^{-d}\sum_{i}\hat{\sigma}_{i}$, the average of which is the magnetization $m=\langle\hat{s}\rangle$. We represent this quantity by $\hat{P}(\hat{s}=s)$. The fluctuations of $\hat{s}$ are measured by the quantity $\langle\hat{s}^{2}\rangle=L^{-d}\chi$ with $\chi$ the magnetic susceptibility. For a weakly correlated system,  i.e.  $T\neq T_{c}$ and in the limit of infinite volume ($L\rightarrow\infty$), one recovers (a generalization of) the Central Limit Theorem, i.e. we find ${\hat{P}(\hat{s}=s)}\propto \exp{\left(-\frac{L^{d}s^2}{2\chi}\right)}$. However, the same fails to hold in case of strongly correlated systems where $\chi$ scales with $L$ as $\chi\propto L^{2-\eta}$ (with $\eta$ the anomalous dimension). This ultimately leads to the failure of CLT in the limit $L\rightarrow\infty$ and at $T=T_c$. Thus what we will be mostly interested in is not just any PDF but the critical PDFs in the thermodynamic limit  (where the thermodynamic limit and the limit of criticality is taken simultaneously).The scaling of the fluctuations of $\hat{s}$ also suggests that for each $\zeta=\xi_\infty/L$, a universal PDF can be built from $\hat{P}(\hat{s}=s)$ once expressed in terms of the scaling variable  $\tilde{s}=sL^{\beta/\nu}$\cite{balog2022critical, eisenriegler1987helmholtz, bouchaud1990anomalous}. The calculation of $\hat{P}(\hat{s}=s)$ is  performed within the functional perturbative $\epsilon$-expansion.\\
Since the PDF is a universal quantity, we work in the continuum with the $\phi^4$ theory and periodic boundary conditions. Thus, the object of interest is:
\begin{equation}\label{P}
    \hat{P}(\hat{s}=s) \propto\int D\hat{\phi}~\delta(\hat{s}-s)\exp(-\int\mathcal{H}[\hat{\phi}]),
\end{equation}
where:
\begin{equation}
    \hat{s}=\frac{1}{L^d}\int_{x}\hat{\phi}(x)
\end{equation}
with $\int_{x}=\int d^{d}x$, and the Hamiltonian
$\mathcal{H}$ is given by:
\begin{equation}\label{Hamil}
    \mathcal{H}[\hat{\phi}]=\int_{x}  \frac{Z_{1}}{2}\left(\nabla\hat{\phi}(x)\right)^2+\frac{1}{2}Z_{2}\,t\,{\hat{\phi}}^2(x)+\frac{1}{4!}Z_{4}\,g\,{\hat{\phi}}^4(x).
\end{equation}
Here we have chosen to parameterize the theory in terms of the renormalized coupling constants $t$, $g$ and renormalized field $\hat{\phi}$ defined at an infrared scale $\mu$ instead of the bare coupling constants: $t_{0}$, $g_{0}$ and bare field $\hat{\phi}_{\Lambda}$ which are defined at a UV scale $\Lambda\sim a^{-1}$ where $a$ is reminiscent of the lattice spacing of the Ising model. The renormalized and the bare couplings and field are related by the following relations :
\begin{equation}
    t_{0}=Z'_{2}t\ , \ \ g_{0}=Z'_{4}g\ , \ \ \phi_{\Lambda}=Z^{1/2}_{1}\hat{\phi},
\end{equation}
with 
\begin{equation}
    Z_{2}=Z_{1}Z'_{2}\ , \ \ Z_{4}=Z^{2}_{1}Z'_{4}\ ,
\end{equation}
 In the following, we use dimensional regularization, where $Z_{1}$, $Z_{2}$ and $Z_{4}$ are counter-terms introduced in accordance with the MS scheme.

Replacing the delta function in Eq.\eqref{P} with an infinitely peaked Gaussian i.e. $\delta(z)\propto\exp\left(-\frac{M^2 z^2}{2}\right)$ with $M\rightarrow\infty$, the PDF can be interpreted as the partition function (with $\mathcal{N}$ a normalization constant):
 \begin{equation}\label{b}
     \mathcal{Z}_{M,s}[h] =\mathcal{N}\int D\hat{\phi}\exp\left(-\mathcal{H}_{M,s}+\int_x h(x)\hat{\phi}(x)\right),
 \end{equation}  
  at vanishing magnetic field $h=0$  of a system with  Hamiltonian
 \begin{equation}
     \mathcal{H}_{M,s}\left(\hat{\phi}(x)\right)=\mathcal{H}\left(\hat{\phi}(x)\right)+ \frac{M^2}{2}\left(\int_{x}(\hat{\phi}(x)-s)\right)^2.
 \end{equation} 
 We also define the modified Legendre transform:
\begin{equation}\label{PE1}
    \Gamma_{M}[\phi]+\log{\mathcal{Z}_{M,s}[h]} = h.\phi-\frac{M^2}{2}\left(\int_{x}({\phi}(x)-s)\right)^2,
\end{equation}
with
\begin{equation}\label{phi}
     \phi(x) = \frac{\delta \log{\mathcal{Z}_{M, s}[h]}}{\delta h(x)}, 
 \end{equation}
 $\left(h.\phi=\int_{x} h(x)\phi(x) \right)$ and $\Gamma_{M}[\phi]$ the modified effective potential.
 We also subsequently define:
 \begin{equation}
     W_{M,s}[h]=\log{\mathcal{Z}_{M,s}[h]}.
 \end{equation}
 Following \cite{sahu2024generalization} and also appendix \ref{appenA}, one can show :
 \begin{align}
     \lim_{M\rightarrow\infty} e^{-\Gamma_{M}[\phi(x)=s]} & = \mathcal{N}\int D\hat{\phi}~\delta\left(\int{\hat\phi(x)}-s \right)e^{-\int \mathcal{H}[\hat{\phi}]}\nonumber\\
     & \propto {\hat{P}(\hat{s}=s)}.
 \end{align}
 We define the quantity ${\hat{I}(s, \xi_{\infty}, L)}$ in the following way:
 \begin{equation}
    {\hat{P}(\hat{s}=s)}\propto e^{-L^{d}{\hat{I}(s,\, \xi_{\infty},\,  L)}},
 \end{equation}
 and thus :
\begin{equation}\label{i(s)}
    \lim_{M\rightarrow\infty}\Gamma_{M}[\phi(x)=s] =L^{d}{\hat{I}(s,\, \xi_{\infty},\,  L)}.
\end{equation}
\section{Renormalized Loop Expansion of the Rate Function}\label{2loopcal}
\subsection{One-Loop Calculation} 
In the following , we briefly review the one-loop calculation. Details are given in \cite{sahu2024generalization} and appendix \ref{appenB}. Expanding $\mathcal{Z}_{M, s}$ around the mean field configuration $\hat{\phi}_{0}$, defined by:
\begin{equation}
    \frac{\delta \mathcal{H}_{M, s}}{\delta \hat{\phi}}|_{\hat{\phi}=\hat{\phi}_{0}}=h.
\end{equation}
One obtains :
\begin{equation}
    \Gamma_{M}[\phi] = \mathcal{H}[\phi]+\frac{1}{2}\Tr\log\mathcal{H}^{(2)}_{M, s}-\log{[\mathcal{N}]},
\end{equation}
where $\log[\mathcal{N}]$ has  been chosen such that it is equal to the value of $\mathcal{H}[\phi]+\Tr\log\mathcal{H}^{(2)}_{M, s}[\phi]$ computed at vanishing field and vanishing mass (i.e. at $T=T_{c}$) with an addition of a constant-infinite pole counter-term to cancel out the divergence coming out from the one-loop term.
\subsection{Two-Loop Calculation}
We start with the two-loop computation of $W_{M,s}[h]$. In the derivation that follows, we use an auxiliary $\hbar$ as means of counting loops. Hence, the full $W_{M,s}[h]$ up to all orders of perturbation theory is given by:
\begin{align}\label{PEN2}
W_{M,s}[h] &=-\mathcal{H}_{M,s}[\hat{\phi}_{0}]+ \int_{x} h(x)\hat{\phi}_{0}(x)-\frac{\hbar}{2}\Tr\log\mathcal{H}^{(2)}_{M, s}[\hat{\phi}_{0}]\nonumber\\
& +W^{(2)}_{M,s}[h]-\log{[\mathcal{N}]},
\end{align}
where $W^{(2)}_{M,s}[h]$ is given by:
\begin{align}
W^{(2)}_{M,s}[h]=\hbar\log{Z^{(2)}_{M,s}[h]},
\end{align}
with 
\begin{align}\label{Z2 M}
& Z^{(2)}_{M,s}[h]\nonumber\\
& =\frac{\int \mathcal{D\hat{\phi}}\exp{-\frac{1}{\hbar}\left(\mathcal{H}_{M,s}[\hat{\phi}+\hat{\phi}_{0}]- \int_{x} h(x)\hat{\phi}_{0}(x)-\mathcal{H}_{M,s}[\hat{\phi}]\right)}}{\int \mathcal{D\hat{\phi}}\exp{-\frac{1}{\hbar}\left(\int\hat{\phi}.\frac{\delta^2\mathcal{H}_{M,s}}{\delta\hat{\phi}^2}|_{\hat{\phi}=\hat{\phi}_{0}}.\hat{\phi}\right)}}
\end{align}
[Here $\log{\mathcal{N}}$ is now modified so as to include constant pole terms of higher orders as well].
Considering only terms up to two-loop order, one must expand Eq.\eqref{Z2 M} up to the fourth order in $\hat{\phi}$. Doing so and proceeding as in Appendix \ref{appenB} one ends up with:
\begin{align}\label{feyn}
    W^{(2)}_{M,s}[h] & = \frac{\hbar^2}{12}\mathcal{I}_{1}+\frac{\hbar^2}{8}\mathcal{I}_{2}-\frac{\hbar^2}{8}\mathcal{I}_{3},
\end{align}
with :\\

\begin{tikzpicture}
    \begin{feynman}
     \vertex(a);
     \vertex [right =of a] (b); 
     \vertex [left=0.3em of a] {$\mathcal{I}_{1}=
$}; 
\vertex [right=0.5em of b];
 \diagram{      
            (a) -- (b); 
            (a) --(b);
            (a) --[half left](b);
            (a) --[half right](b);
            
        };

    \end{feynman}
\end{tikzpicture}\\

 \begin{tikzpicture}[scale=0.3]
    \begin{feynman}
     \vertex(a); 
     \vertex[left=0.5em of a]{$\mathcal{I}_{2}=$};
     \vertex [right =of a] (b); 
      \vertex [above =of a] (c);
      \vertex [above =of b] (d);
 \diagram{      

            (a) -- [quarter left](c);
            (a) -- [quarter right](c);
            (a) -- (b);
            (b) -- [quarter left](d);
            (b) -- [quarter right](d);

        };

    \end{feynman}
    \end{tikzpicture}\\
    
\begin{tikzpicture}

    \begin{feynman}
     \vertex(a);
     \vertex [right =of a] (b); 
      \vertex [right =of b] (c);
      \vertex [left=0.2em of a] {$\mathcal{I}_{3}=
$};
      
        \diagram{
            
            (a) [scalar]--[half right](b);
            (a) --[half left](b);
            (b) --[half left](c);
            (b) --[half right](c);
        };

    \end{feynman}
    
   \end{tikzpicture}\\
The propagator is as usual given by:
\begin{equation}\label{D2 M}
    \mathcal{D}^{-1}=\frac{\delta^2 \mathcal{H}_{M, s}}{\delta \phi^2}.
\end{equation}
All the diagrams are hence computed using this `modified propagator'. This marks the difference between the effective potential and the rate functions. Another very important comment is in order. From \cite{jackiw1974functional}, one can also show that for obtaining the effective potential of the $\phi^4$ theory, it just suffices to compute the 1-Particle-Irreducible (1PI) diagrams generated by $\frac{g\phi}{6}\hat{\phi}^3+\frac{g}{24}\hat{\phi}^4$. Using similar arguments, one can also show that for computing the rate function and consequently the PDF at higher orders of perturbation theory, one needs to compute the 1PI-s generated by $\frac{g\phi}{6}\hat{\phi}^3+\frac{g}{24}\hat{\phi}^4 $ with the `modified propagator', taking the limit of $M\rightarrow\infty$ at the end of the calculation. Thus, in principle we should be able to perturbatively compute the PDF at any order using the current framework. However this is beyond the scope of the present work and to be fair is a seemingly daunting task starting from three loops.\\

Using Eq.\eqref{PEN2} and expressing $\hat{\phi}_{0}$ in terms of $\phi$ following Appendix \ref{appenB}, we obtain :
\begin{align}\label{GammaP M}
\Gamma_{M}[\phi] & =  \mathcal{H}[\phi]+\frac{\hbar}{2}\Tr\log{\mathcal{H}^{(2)}_{M,s}}-W^{(2)}_{M,s}[h]-\log{[\mathcal{N}]}\nonumber\\
& +\frac{\hbar^2}{8}\Tr\left(\mathcal{D}\frac{\delta}{\delta \phi}\mathcal{D}^{-1}\right).\mathcal{D}.\Tr\left(\mathcal{D}\frac{\delta}{\delta \phi}\mathcal{D}^{-1}\right).
\end{align}
Using Eqs.~\eqref{GammaP M} and \eqref{feyn}, at two-loop order one thus ends up with (where the $\phi$ dependence of $\Gamma_{M}$ has been omitted):
\begin{equation}\label{Gammarate M}
 \Gamma_{M}  =  \mathcal{H}+\frac{\hbar}{2}\Tr\log{\mathcal{H}^{(2)}_{M,s}}\\
 +\frac{\hbar^2}{8}\mathcal{I}_{3}-\frac{\hbar^2}{12}\mathcal{I}_{1}-\log[\mathcal{N}].
 \end{equation}
 One important thing to notice here is that the propagator of this theory is a bit different from the usual $\phi^4$ theory in the continuum since it is given by Eq.~\eqref{D2 M}. The propagator in Fourier space for the $\phi^{4}$ theory as parameterized in Eq.~\eqref{Hamil} is hence given by :
 \begin{figure}[h]
 \begin{tikzpicture}[scale=0.4]
    \begin{feynman}
     \vertex(a);
     \vertex [right =of a] (b); 
     \vertex [left=0.3em of a] {$\mathcal{D}^{-1}=
$}; 
\vertex [right=0.4em of b] {$= Z_{1}q^2+M^2\delta_{q, 0}+Z_{2}t+Z_{4}{g\phi ^2}/{2}.
$};
 \diagram{      
            (a) -- (b); 
            
        };

    \end{feynman}
\end{tikzpicture} 
 \end{figure}
  
Using $\mathcal{D}^{-1}=Z_{1}q^2+M^2\delta_{q, 0}+Z_{2}t+Z_{4}{g\phi ^2}/{2}$, one thus obtains for a $\phi^{4}$ theory :
\begin{widetext}
\begin{align}\label{e}
      \Gamma_M[\phi]& = L^d\biggl(\frac{1}{2}Z_{2}t \phi^2+\frac{1}{4!}Z_{4}g \phi^4+\hbar\frac{1}{2L^d}\sum_{\vec{q}}\log\left(1+\frac{m^2}{Z_{1}\vec{q}^2+M^2\delta_{\vec{q}, 0}}\right)+\frac{g}{8}\hbar^2\left(\frac{1}{L^{d}}\sum_{\vec{q}}\frac{1}{Z_{1}\vec{q}^2+m^2+M^2\delta_{q, 0}}\right)^2\nonumber\\
 &-\frac{g^2\phi^2}{12}\hbar^2\frac{1}{L^{2d}}\sum_{\vec{p}, \vec{q}}\frac{1}{(Z_{1}\vec{p}^2+m^2+M^2\delta_{\vec{p}, 0})(Z_{1}\vec{q}^2+m^2+M^2\delta_{\vec{q}, 0})(Z_{1}(\vec{p}+\vec{q})^2+m^2+M^2\delta_{\vec{p}+\vec{q}, 0})}\biggr)-\log{[\mathcal{N}']},
  \end{align}
\end{widetext}
with $m^2 = Z_{2}t+Z_{4}{g\phi ^2}/{2}$ and $\log{\left[\mathcal{N}\right]}-\log{\left[\mathcal{N}'\right]}=\mathcal{H}[\phi]|_{\phi=0}+\frac{\hbar}{2}\Tr\log{\mathcal{H}^{(2)}_{M,s}}[\phi]|_{\phi=0}$.\\
Since the field renormalization $(Z_{1}-1)$ is itself of the order $\hbar^2$, it doesn't directly play any role in the computation of the rate function at two loops. [In the limit of criticality, the coupling constant in front of the $\phi^4$ term flows to its fixed point value, the determination of which in $d=4-\epsilon$ is achieved by the explicit computation of $Z_{1}$.] Thus, in the limit $M\rightarrow\infty$,
Eq.\eqref{e} becomes:
\begin{widetext}
 \begin{align}\label{e1}
      \lim_{M\rightarrow\infty}\Gamma_M[\phi(x)=s]& = L^d\biggl(\frac{1}{2}Z_{2}t s^2+\frac{1}{4!}Z_{4}g s^4+\hbar\frac{1}{2L^d}\sum_{\vec{q}\neq 0}\log\left(1+\frac{m^2}{\vec{q}^2}\right)+\frac{g}{8}\hbar^2\left(\frac{1}{L^d}\sum_{\vec{q}\neq 0}\frac{1}{\vec{q}^2+m^2}\right)^2\nonumber\\
 &-\frac{g^2 s^2}{12}\hbar^2\frac{1}{L^{2d}}\sum_{\substack{\{\vec{p},\vec{q}\}\neq 0,\\ \vec{p}\neq -\vec{q}}}\frac{1}{(\vec{p}^2+m^2)(\vec{q}^2+m^2)((\vec{p}+\vec{q})^2+m^2)}\biggr)-\log{[\mathcal{N}']},
  \end{align}   
  with $m^2 = Z_{2}t+Z_{4}{g s^2}/{2}$ .
\end{widetext}
The function $\Gamma_M[\phi(x)=s]$ can be equivalently expressed in terms of dimensionless quantities. $\Gamma_M[\phi(x)=s]$ is a priori defined at an energy scale $\mu$. Thus at this scale $\mu$, the dimensionless renormalized parameters are defined as:
\begin{equation} \label{dimensionless}
\begin{split}
\bar{g}=&\mu^{-\epsilon}g =\mu^{-\epsilon}16\pi^2 \bar{u},\\
 \bar{s}=&\mu^{-1+\epsilon/2}Z^{-1/2}_{1}s,\\
\bar{t}=&\mu^{-2}t,\\
\bar{L}=&\mu L,
\end{split}
\end{equation}
where $Z_{1}=Z_{1}(\bar{u})$ is the field renormalization. Since fluctuations above a scale larger than the system size are forbidden, it is natural to choose the scale $\mu=L^{-1}$, so as to include all the fluctuations between the UV scale $\Lambda$ and the infrared scale $\mu$. At the scale $L^{-1}$, we define the dimensionless variables the following way:
\begin{equation}\label{stilde}
    \tilde{s}=sL^{\frac{d-2+\eta}{2}}\ , \ \tilde{g}=L^{4-d}{g}(\mu=L^{-1})\ , \tilde{t} = L^2 t(\mu = L^{-1}).   
\end{equation}

 In the MS scheme,  $Z_{2}$, $Z_{4}$ and $\log{[\mathcal{N}']}$ at two loops are given by \cite{kleinert2001critical}:
\begin{equation}\label{counterterms}
 \begin{split}
    Z_{2} & = 1 + \frac{\tilde{g} \hbar}{16\pi^2\epsilon}+\frac{2\tilde{g}^{2}\hbar^2}{(16\pi^2)^2\epsilon^2}-\frac{\tilde{g}^2\hbar^2}{2(16\pi^2)^2\epsilon},\\
    Z_{4} & = 1 + \frac{3\tilde{g} \hbar}{16\pi^2\epsilon}+\frac{9\tilde{g}^{2}\hbar^2}{(16\pi^2)^2\epsilon^2}-\frac{3\tilde{ g}^2\hbar^2}{(16\pi^2)^2\epsilon},\\
    \log{\left[\mathcal{N}'\right]}&=\tilde{t}^{2}\left(\frac{ \hbar}{2(16\pi^{2})\epsilon}+\frac{\tilde{g}\hbar^{2}}{2(16\pi^2)^2\epsilon^{2}}\right).
 \end{split}
\end{equation}
In Appendix \ref{appen C}, we have shown how to compute the diagrams emerging at two-loops. Using the one-loop result from \cite{sahu2024generalization} and Eqs.~\eqref{counterterms}, \eqref{div-rate2} and \eqref{conv-rate2} one thus obtains (A detailed calculation of the quantity $\lim_{M\rightarrow\infty}\Gamma_M[\tilde{s}, L^{-1}]$ has been given in Appendix \ref{appen D} of this paper):
\begin{align}\label{convG}
\lim_{M\rightarrow\infty}\Gamma_M[\tilde{s}, L^{-1}]=\Gamma^{(div)}[\tilde{s},L^{-1}]+\Gamma^{(conv)}[\tilde{s},L^{-1}],
\end{align}
with
\begin{widetext}
 \begin{align}\label{divc-rate}
  \Gamma^{(div)}[\tilde{s},L^{-1}]   & = \frac{1}{2}\left(\frac{\tilde{g} \hbar}{16\pi^2\epsilon}+\frac{2\tilde{g}^{2}\hbar^2}{(16\pi^2)^2\epsilon^2}-\frac{\tilde{g}^2\hbar^2}{2(16\pi^2)^2\epsilon}\right)\tilde{t}\tilde{s}^{2}+\frac{1}{24}\left(\frac{3\tilde{g} \hbar}{16\pi^2\epsilon}+\frac{9\tilde{g}^{2}\hbar^2}{(16\pi^2)^2\epsilon^2}-\frac{3\tilde{ g}^2\hbar^2}{(16\pi^2)^2\epsilon}\right)\tilde{g}\tilde{s}^{4}\nonumber\\
  & -\frac{\hbar}{2(16\pi^2)\epsilon}\left(\tilde{t}+\frac{\tilde{g}\tilde{s}^2}{2}\right)^{2}-\hbar^{2}\frac{\tilde{g}^{2}\tilde{s}^2}{12} \frac{\left(\tilde{t}+\tilde{g}\tilde{s}^2/2\right)}{(16\pi^2)^2}\left[-\frac{6}{\epsilon^2} +\frac{6}{\epsilon}\left\{\gamma_{E}-\frac{3}{2}+\log{\left(\frac{\left(\tilde{t}+\tilde{g}\tilde{s}^2/2\right)}{4\pi}\right)}\right\}\right]\nonumber\\
  & +\hbar^{2}\frac{\tilde{g}}{8}\frac{\left(\tilde{t}+\tilde{g}\tilde{s}^{2}/2\right)^{2}}{(16\pi^2)^2}\left[\frac{4}{\epsilon^{2}}-\frac{4}{\epsilon}\left\{\gamma_{E} -1
     +\log{\left(\frac{\left(\tilde{t}+\tilde{g}\tilde{s}^2/2\right)}{4\pi}\right)}\right\}\right]-\hbar^{2}\frac{\tilde{g}^{2}\tilde{s}^{2}}{32\pi^{2}\epsilon}\left\{\frac{1}{4\pi}\theta^{(1)}\left(\frac{\left(\tilde{t}+\tilde{g}\tilde{s}^2/2\right)}{4\pi}\right)\right\}\nonumber\\
     & -\hbar^{2}\frac{\tilde{g}}{32\pi^{2}\epsilon}\left(\tilde{t}+\frac{\tilde{g}\tilde{s}^{2}}{2}\right)\left\{\frac{1}{4\pi}\theta^{(1)}\left(\frac{\left(\tilde{t}+\tilde{g}\tilde{s}^2/2\right)}{4\pi}\right)\right\}+\frac{\hbar^2}{2(16\pi^2)^2\epsilon}\tilde{g}\left(\tilde{t}+\frac{3\tilde{g}\tilde{s}^2}{2}\right)\left(\tilde{t}+\frac{\tilde{g}\tilde{s}^2}{2}\right)\biggl[-\frac{2}{\epsilon}+\gamma_{E}-1\nonumber\\
       & +\log{\left(\frac{\tilde{t}+\tilde{g}\tilde{s}^2/2}{4\pi}\right)}\biggr]+\hbar^{2}\frac{\tilde{g}}{32\pi^{2}\epsilon}\left(\tilde{t}+\frac{3\tilde{g}\tilde{s}^{2}}{2}\right)\left\{\frac{1}{4\pi}\theta^{(1)}\left(\frac{\left(\tilde{t}+\tilde{g}\tilde{s}^2/2\right)}{4\pi}\right)\right\}-\tilde{t}^{2}\left(\frac{ \hbar}{2(16\pi^{2})\epsilon}+\frac{\tilde{g}\hbar^{2}}{2(16\pi^2)^2\epsilon^{2}}\right),
 \end{align}   
 and
\begin{align}\label{conv-rate}
   \Gamma^{(conv)}[\tilde{s},L^{-1}] & =   \frac{1}{2}\tilde{t}\tilde{s}^{2}+  \frac{1}{24}\tilde{g}\tilde{s}^{4}+\frac{\hbar}{4(16\pi^2)}\left(\tilde{t}+\frac{\tilde{g}\tilde{s}^2}{2}\right)^{2} \left[\gamma_{E}-\frac{3}{2}+\log{\left(\frac{\tilde{t}+\tilde{g}\tilde{s}^2/2}{4\pi}\right)}\right] +\frac{\hbar}{2}\Delta\left(\frac{1}{4\pi}\left[\tilde{t}+\frac{\tilde{g}\tilde{s}^2}{2}\right]\right)\nonumber\\
   &  -\hbar^{2}\frac{\tilde{g}^{2}\tilde{s}^2}{12} \frac{\left(\tilde{t}+\tilde{g}\tilde{s}^2/2\right)}{(16\pi^2)^2}\left[-3 A-\frac{21}{2}+9\gamma_{E}-3\gamma_{E}^2-\frac{\pi^2}{4}-(6\gamma_{E}-9)\log{\left(\frac{\left(\tilde{t}+\tilde{g}\tilde{s}^2/2\right)}{4\pi}\right)}-3\log^{2}{\left(\frac{\left(\tilde{t}+\tilde{g}\tilde{s}^2/2\right)}{4\pi}\right)}\right]\nonumber\\
  & +\hbar^2\frac{\tilde{g}}{8}\left[\frac{\left(\tilde{t}+\tilde{g}\tilde{s}^2/2\right)^2}{(16\pi^2)^2}\left(2\gamma_{E}^2-4\gamma_{E}+3+\frac{\pi^2}{6}+4(\gamma_{E}-1)\log{\left(\frac{\left(\tilde{t}+\tilde{g}\tilde{s}^2/2\right)}{4\pi}\right)}+2\log^{2}{\left(\frac{\tilde{t}+\tilde{g}\tilde{s}^2/2}{4\pi}\right)}\right)\right]\nonumber\\
  & -\frac{\hbar^2}{8(16\pi^2)^2}\tilde{g}\left(\tilde{t}+\frac{3\tilde{g}\tilde{s}^2}{2}\right)\left(\tilde{t}+\frac{\tilde{g}\tilde{s}^2}{2}\right)\left(\frac{\pi^2}{6}+2-2\gamma_{E}+\gamma_{E}^2+2(\gamma_{E}-1)\log{\left(\frac{ \tilde{t}+\tilde{g}\tilde{s}^2/2}{4\pi}\right)}+\log^{2}{\left(\frac{ \tilde{t}+\tilde{g}\tilde{s}^2/2}{4\pi}\right)}\right)\nonumber\\ 
 & -\hbar^{2}\frac{\tilde{g}^{2}\tilde{s}^2}{12}\biggl[ I_{1}[\tilde{t}+\tilde{g}\tilde{s}^2/2]+I_{2}[\tilde{t}+\tilde{g}\tilde{s}^2/2]+\frac{3}{16\pi^2}\left(\log{4\pi}-\gamma_{E}-\log{\left(\tilde{t}+\tilde{g}\tilde{s}^2/2\right)}\right)\frac{1}{4\pi}\theta^{(1)}\left(\frac{\left(\tilde{t}+\tilde{g}\tilde{s}^2/2\right)}{4\pi}\right)\nonumber\\
   &+\frac{3}{16\pi^2}\frac{1}{\left(\tilde{t}+\tilde{g}\tilde{s}^2/2\right)} \theta^{(2)}\left(\frac{\left(\tilde{t}+\tilde{g}\tilde{s}^2/2\right)}{4\pi}\right)-\frac{1}{\left(\tilde{t}+\tilde{g}\tilde{s}^2/2\right)^3}\biggr]+\hbar^2\frac{\tilde{g}}{8}\biggl[\frac{1}{16\pi^2}\left[\theta^{(1)}\left(\frac{\left(\tilde{t}+\tilde{g}\tilde{s}^2/2\right)}{4\pi}\right)\right]^2\nonumber\\
    & +\frac{2}{4\pi}\theta^{(1)}\left(\frac{\left(\tilde{t}+\tilde{g}\tilde{s}^2/2\right)}{4\pi}\right)\frac{\left(\tilde{t}+\tilde{g}\tilde{s}^2/2\right)}{(4\pi)^2}\left\{\gamma_{E} -1+\log{\left(\frac{\left(\tilde{t}+\tilde{g}\tilde{s}^2/2\right)}{4\pi}\right)}\right\}\biggr]-\frac{\epsilon\hbar}{2}\Delta^{(\epsilon)}\left(\frac{1}{4\pi}\left[\tilde{t}+\frac{\tilde{g}\tilde{s}^2}{2}\right]\right)\nonumber\\
    & -\frac{\epsilon\hbar}{2}\frac{1}{48}\left(\frac{\tilde{t}+\tilde{g}\tilde{s}^2/2}{4\pi}\right)^2\left[21- 18\gamma_{E} + 6 \gamma_{E}^2 + \pi^2 + 6 (2\gamma_{E}-3 )\log{\left(\frac{\tilde{t}+\tilde{g}\tilde{s}^2/2}{4\pi}\right)}+6 \log^{2}{\left(\frac{\tilde{t}+\tilde{g}\tilde{s}^2/2}{4\pi}\right)} \right].
   \end{align}
    Here $\gamma_{E}$ is the Euler-Mascheroni constant. The functions $\theta(\tilde{m}^{2})$, $\theta^{(1)}(\tilde{m}^{2})$, $\theta^{(2)}(\tilde{m}^{2})$, and the constant $A$ are the same as defined in Appendix \ref{appen C} and the functions $I_{1}[\tilde{m}^2]$, $I_{2}[\tilde{m}^2]$, $\Delta(\tilde{m}^2)$, $\Delta^{(\epsilon)}(\tilde{m}^2)$ are the same as defined in the Appendix \ref{appen D} of this paper. The integrals $I_{1}[\tilde{m}^{2}]$ and $I_{2}[\tilde{m}^{2}]$ are rather demanding to compute. Numerical integration of these integrals follow from methods shown in \cite{bijnens2014two}.
\end{widetext}
As is evident from Eqs.\eqref{divc-rate} and \eqref{conv-rate}, $\Gamma^{(conv)}[\tilde{s}, L^{-1}]$ is {\it a priori} convergent and is devoid of any poles in $\epsilon$ while $\Gamma^{(div)}[\tilde{s}, L^{-1}]$ contains all the poles in $\epsilon$ and is thereby divergent for $\epsilon\rightarrow 0$ i.e. at $d=4$  . After some elementary calculations, one can show that all the pole terms in $\epsilon$ i.e. terms of the order $\frac{1}{\epsilon}$ and $\frac{1}{\epsilon^{2}}$ actually cancel out, i.e. $\Gamma^{(div)}[\tilde{s}, L^{-1}]=0$. Hence, what we are left with in the end is a theory free of any divergence or $\lim_{M\rightarrow\infty}\Gamma_{M}[\tilde{s}, L^{-1}]=\Gamma^{(conv)}[\tilde{s}, L^{-1}]$. At this stage, another important comment is also in order. Broadly speaking, the divergences in $\Gamma^{(div)}[\tilde{s}, L^{-1}]$ can be divided into two types: (i) local and (ii) non-local. The local divergences are by and large harmless, since the counter-terms are introduced specifically in a way to cancel out these local divergences. The non-local divergences on the other hand should cancel each other out order by order in the perturbation theory. The counter-terms introduced in Eq.\eqref{counterterms} are the same as for a theory in the infinite-volume limit. There arises extra non-local divergences due to the finite-size corrections. The non-local divergence associated with the finite-size corrections is given by : $-\frac{\tilde{g}}{32\pi^2\epsilon}\left(\tilde{t}+\frac{3\tilde{g}\tilde{s}^2}{2}\right)\frac{1}{4\pi}\theta^{(1)}\left(\frac{\tilde{t}+\tilde{g}\tilde{s}^2/2}{4\pi}\right)$. This term is exactly cancelled out as is seen in Eq.\eqref{divc-rate}, thereby confirming the perturbative renormalizability of the theory. This also means that the theory in both finite and infinite volume has exactly the same kind of UV divergences which also fits our intuition.

Universality takes place in the double limit of infinite volume, i.e. $L\rightarrow\infty$ and criticality $t\rightarrow 0$. This renormalized temperature (difference) $t$  is linked to the correlation length of the system as $\xi_{\infty} \sim t^{-1/\nu}$ asymptotically close to criticality. As shown in the previous one-loop calculation and in \cite{balog2022critical, ranccon2025probability}, the double limit of criticality and infinite volume is not unique and may be approached in various ways keeping $\zeta=\lim_{L,\xi_{\infty}\rightarrow\infty}\frac{L}{\xi_{\infty}}$ fixed. Thus, when $L\to\infty$ and $T\rightarrow T_{c}$, we end up with a family of PDFs and rate functions:
\begin{equation}\label{pdf34}
\begin{split}
  \hat{P}(\hat{s}=s, T, L) & \propto\, e^{-L^{d}\hat{I}(s, \xi_{\infty}, L)}\\
  &=L^{\beta/\nu}P_{\zeta}(\tilde{s})\propto e^{-I_{\zeta}(\tilde{s})}.
\end{split}    
\end{equation}
where $P_\zeta$ and $I_\zeta$ are universal functions of $\tilde s$ \footnote{It is important to note here that the functions $\hat{I}(s, \xi_{\infty}, L)$ and $\hat{P}(\hat{s}=s, T, L)$ are not universal but the family of functions $I_{\zeta}(\tilde{s})$ and consequently $P_{\zeta}(\tilde{s})$  are.}\\

If we were to run a RG flow, then we know that at the scale $\mu=L^{-1}$ with $L\rightarrow\infty$, the dimensionless coupling constant $\tilde{g}$ would flow to its fixed point value $\tilde{g}=16\pi^2 u_{*}$ with $u_{*}=\frac{\epsilon}{3}+\frac{17}{81}\epsilon^{2}$ for the Ising model (at two loops). For an infrared scale $\mu\ll\Lambda$, the correlation length is given by $\xi_{\infty}^{-1}=\mu\left(\frac{t}{\mu^2}\right)^{\nu}$. Since, we have chosen to define the renormalized parameters at the scale $\mu=L^{-1}$, we end up with the relation $tL^{2}=\tilde{t}=\zeta^{1/\nu}$ . Defining $x=\sqrt{u_{*}}\tilde{s}$ and following Eq.\eqref{conv-rate}, the full two-loop expression for the rate function $I_{\zeta}(x)$ is hence given as (factors of $\hbar$ are now omitted):
\begin{widetext}
    \begin{align}\label{ratef}
        I_{\zeta}(x) & = \frac{1}{u_{*}}\biggl[\frac{1}{2}\zeta^{1/\nu}x^{2}+\frac{2\pi^2}{3}x^{4}+\epsilon\left\{\frac{\pi^2}{12}\left(\frac{\zeta^{1/\nu}}{4\pi^2}+2x^{2}\right)^{2}\left(\gamma_{E}+\log{2\pi}-\frac{3}{2}+\log{\left(\frac{\zeta^{1/\nu}}{8\pi^2}+x^{2}\right)}\right)+\frac{1}{6}\Delta\left(\frac{\zeta^{1/\nu}+8\pi^{2}{x}^2}{4\pi}\right)\right\}\nonumber\\
        &+\epsilon^{2}\Bigg\{-\frac{1}{41472\pi^{2}}\left(\zeta^{1/\nu}+8\pi^{2}x^{2}\right)\biggl(72\pi^{4}x^{2}+(357-226\gamma_{E}+18\gamma_{E}^{2})\zeta^{1/\nu}+\pi^{2}\left(-8[3+144 A -62\gamma_{E}+54\gamma_{E}^{2}]x^{2}+9\zeta^{1/\nu}\right)\nonumber\\
        &+\left(16(31-54\gamma_{E})\pi^{2}x^{2}+2(18\gamma_{E}-113)\zeta^{1/\nu}\right)\log{\left(\frac{\zeta^{1/\nu}+8\pi^{2}x^{2}}{4\pi}\right)}+18\left(\zeta^{1/\nu}-24\pi^{2}x^{2}\right)\log^{2}{\left(\frac{\zeta^{1/\nu}+8\pi^{2}x^{2}}{4\pi}\right)}\biggr)\nonumber\\
       &-\frac{(16\pi^{2})^{2}x^{2}}{108}\biggl[I_{1}\left[\zeta^{1/\nu}+8\pi^{2}{x}^2\right]+I_{2}\left[\zeta^{1/\nu}+8\pi^{2}{x}^2\right]+\frac{3}{16\pi^{2}\left(\zeta^{1/\nu}+8\pi^{2}{x}^2\right)}\theta^{(2)}\left(\frac{\zeta^{1/\nu}+8\pi^{2}{x}^2}{4\pi}\right)\nonumber\\
        & +\frac{3}{16\pi^2}\left(\log{4\pi}-\gamma_{E}-\log{\left(\zeta^{1/\nu}+8\pi^{2}{x}^2\right)}\right)\frac{1}{4\pi}\theta^{(1)}\left(\frac{\zeta^{1/\nu}+8\pi^{2}{x}^2}{4\pi}\right)-\frac{1}{(\zeta^{1/\nu}+8\pi^{2}x^{2})^{3}}\biggr]\nonumber\\
        &+\frac{2\pi^{2}}{9}\biggl[\frac{1}{16\pi^2}\left(\theta^{(1)}\left[\left(\frac{\zeta^{1/\nu}+8\pi^{2}x^{2}}{4\pi}\right)\right]\right)^2 +\frac{2}{4\pi}\theta^{(1)}\left(\frac{\zeta^{1/\nu}+8\pi^{2}x^{2}}{4\pi}\right)\left(\frac{\zeta^{1/\nu}+8\pi^{2}x^{2}}{(4\pi)^2}\right)\left(\gamma_{E} -1+\log{\left(\frac{\zeta^{1/\nu}+8\pi^{2}x^{2}}{4\pi}\right)}\right)\biggr]\nonumber\\
        &+\frac{17}{162}\Delta\left(\frac{\zeta^{1/\nu}+8\pi^{2}{x}^2}{4\pi}\right)-\frac{1}{6}\Delta^{(\epsilon)}\left(\frac{\zeta^{1/\nu}+8\pi^{2}{x}^2}{4\pi}\right)\Bigg\}\biggr].
    \end{align}
\end{widetext}
We stress here the fact that using the variable $x$ rather than $\tilde{s}$ is on one hand a finite redefinition of the field when $\epsilon$ is fixed and non-vanishing and on the other hand guarantees a systematic $\epsilon$ expansion. We also note that the $\epsilon\rightarrow 0$ limit of the theory is ill-defined and hence the limit of $d=4$ cannot be studied using the current framework. However this is not really a surprise as explained in \cite{lawrie1976field}.\\

A direct consequence of Eq.~\eqref{i(s)} is that the quantity $ L^{d}\hat{I}(s,\xi_{\infty}, L)$ is nothing but the free energy of a system with the modified Hamiltonian $\mathcal{H}_{M, s}$. Using standard RG arguments, one can readily show that the free energy of a system near criticality is a scaling function when properly expressed in terms of scaling variables. Eq.\eqref{ratef} is thereby a perturbative demonstration of the fact that the quantity $ L^{d}\hat{I}(s,\xi_{\infty}, L)$ is indeed just a function of the scaling variables $\tilde{s}$  and $\zeta$ (remember $x$ is nothing but a finite rescaling of $\tilde{s}$ with a $\sqrt{u_{*}}$).\\

The derivation of Eq.~\eqref{ratef} was made possible due to the natural choice of infrared scale, i.e. $\mu=L^{-1}$ at which the renormalized parameters of the theory were defined. However, this derivation could also be done by defining the renormalized parameters of the theory at any arbitrary scale $\mu$. In the following we give a hint of how to do it. Defining $\delta \gamma_{M}$ as the free energy density of a system with free energy $\Gamma_{M}$, i.e.:
\begin{equation}
    \frac{\Gamma_{M}(s, T, L)}{(L\mu)^{d}}=\delta \gamma_{M}(s, T, g, L),
\end{equation}
(with $t$, $g$, $s$, $L$ defined as in \eqref{dimensionless}), one obtains in the $M\rightarrow\infty$ limit:
\begin{equation}\label{SI}
    \mu^{d}\delta\gamma_{\infty}(\bar{s}, \bar{t}, \bar{g}, \bar{L})={\mu'}^{d}\delta\gamma_{\infty}(\bar{s'}, \bar{t'}, \bar{g'}, \bar{L'}),
\end{equation}
where the primed variables are defined at a scale $\mu'$. Defining $l=\frac{\mu'}{\mu}$, the evolution of couplings and fields with $l$ is given by their RG flow. In the double limit of infinite volume $\bar{L}\rightarrow\infty$ and criticality $t\rightarrow 0$, we can choose $l$ to be much smaller than the energy scales at play such that $\bar{g}$ runs to its fixed point value $g_{*}=16\pi^2 u_{*}$ and $Z_{1}(u_{*})\sim \mu^{-\eta}$ with $\eta$ the anomalous dimension. Choosing $\bar{L}=l^{-1}$, one thus obtains:
\begin{align}\label{Sc}
   \delta\gamma_{\infty}(\bar{s}, \bar{t}, \bar{g}, \bar{L}) & = l^{d} \delta\gamma_{\infty}(\bar{s} l^{-\beta/\nu}, \bar{t}l^{-1/\nu}, u_{*}, \bar{L}l)\nonumber\\
   & = \bar{L}^{-d}\delta\gamma_{\infty}(\tilde{s}, \zeta^{1/\nu}, u_{*}, 1),
\end{align}
with $\tilde{s}$ and $\zeta$ defined as before and $\beta/\nu =(d-2+\eta)/2$. Thus, Eq.\eqref{Sc} also shows that the rate function is a scaling function of just two variables $\tilde{s}$ and $\zeta$. This in turn means that the rate function $I_{\zeta}(x)$ can be written as $I_{\zeta}(x)= f(\zeta^{1/\nu}, \bar{x}\bar{L}^{\beta/\nu})$. Since $f(x)$ is itself a scaling function, up-to second order in $\epsilon$ it can be expanded as:
\begin{equation}
    f(x)=f_{0}(x)+\epsilon f_{1}(x)+\epsilon^{2}f_{2}(x).
\end{equation}
Now, expanding $1/\nu$ and $\beta/\nu$ up-to second order in $\epsilon$, the function $f(\zeta^{1/\nu}, \bar{x}\bar{L}^{\beta/\nu})$ is given by:
\begin{widetext}
\begin{align}\label{Sc2}
  f(\zeta^{2-\frac{\epsilon}{3}-\frac{19}{162}\epsilon^{2}}, \bar{x}\bar{L}^{1-\frac{\epsilon}{2}+\frac{\epsilon^{2}}{108}})= & f_{0}\left(\zeta ^2,\bar{x}\bar{L}\right)+{\epsilon}
   \left(-\frac{1}{3} \zeta ^2 \log (\zeta ) f_{0}^{(1,0)}\left(\zeta ^2,\bar{x}\bar{L}\right)-\frac{1}{2} \bar{x}\bar{L} \log (\bar{L}) f_{0}^{(0,1)}\left(\zeta ^2,\bar{x}\bar{L}\right)+f_{1}\left(\zeta ^2,\bar{x}\bar{L}\right)\right)\nonumber\\
   & + {\epsilon}^2 \biggl(\frac{1}{8} (\bar{x}\bar{L})^{2} \log ^2(\bar{L}) f_{0}^{(0,2)}\left(\zeta ^2,\bar{x}\bar{L}\right)+\frac{1}{162} \zeta ^2 \left(9 \log ^2(\zeta )-19 \log (\zeta )\right)
   f_{0}^{(1,0)}\left(\zeta ^2,\bar{x}\bar{L}\right)\nonumber\\
   &+\frac{1}{216} \left(\bar{x}\bar{L}\right) \left(27 \log ^2\bar{L}+2 \log
   \bar{L}\right) f_{0}^{(0,1)}\left(\zeta ^2,\bar{x}\bar{L}\right)+\frac{1}{6} \left(\bar{x}\bar{L}\right) \zeta ^2 \log
   {\zeta } \log \left({\bar{L}}\right) f_{0}^{(1,1)}\left(\zeta ^2,\bar{x}\bar{L}\right)\nonumber\\
   & +\frac{1}{18} f_{0}^{(2,0)}\left(\zeta ^2,\bar{x}\bar{L}\right) \zeta ^4 \log
   ^2{\zeta }-\frac{1}{3}
  f_{1}^{(1,0)}\left(\zeta ^2,\bar{x}\bar{L}\right) \zeta ^2 \log {\zeta}-\frac{1}{2} \left(\bar{x}\bar{L}\right) \log {\left(\bar{L}\right)}
  f_{1}^{(0,1)}\left(\zeta ^2,\bar{x}\bar{L}\right)\nonumber\\
  & +f_{2}\left(\zeta ^2,\bar{x}\bar{L}\right)\biggr), 
\end{align}
\end{widetext}
with:
\begin{align*}
    f^{(n,m)}_{i}(\zeta^{2}, \bar{x}\bar{L})=\left(\frac{\partial}{\partial\zeta^{2}}\right)^{n}\left(\frac{\partial}{\partial(\bar{x}\bar{L})}\right)^{m}f_{i}\left(\zeta^{2}, \bar{x}\bar{L}\right).
\end{align*}\\

Using techniques shown in Appendix \ref{appen C}, one can once again compute $\lim_{M\rightarrow\infty}\Gamma_{M}[\bar{s}, \bar{t}, \bar{L}]$ in $\epsilon-$expansion and identify order by order the functions $f_{0}$, $f_{1}$ and $f_{2}$ (remember $\bar{s}$ and $\bar{x}$ are related by a finite rescaling) from Eq.\eqref{Sc2}. Once again, one obtains the same result as Eq.\eqref{ratef}. The same happens at one-loop as shown in \cite{sahu2024generalization}. Thus, choosing a suitable scale to define the renormalized parameters automatically constructs the scaling function.

\subsection{The Large Field Behaviour}
Using the results derived in the previous section, one could also predict the large field behaviour of the rate function. As shown previously, the rate function itself is a modified effective potential i.e. effective potential in the continuum with finite size corrections. Thus the leading large field behaviour of the rate function is the same as that of the effective potential of a $\phi^{4}$ theory in the continuum and is thus given by $x^{\delta+1}$. In Eq.\eqref{ratef}, we see that in the limit $x\rightarrow\infty$, the leading behaviour of $I_{\zeta}(x)$ is given by:
\begin{align}
I_{\zeta}(x)\sim_{x\rightarrow\infty} \frac{x^{4}}{u_{*}}  \biggl(1+\epsilon\log{x}+\frac{\epsilon^{2}}{54}(25\log{x}+27\log^{2}x)\biggr),
\end{align}
which is nothing but $x^{\delta+1}$ expanded in to second order in $\epsilon$ with $\delta=3+\epsilon+\frac{25}{54}\epsilon^{2}$. However, due to the finite size corrections, the rate function $I_\zeta(x)$ also receives sub-leading coerrections in the tails i.e. in the limit of large field. Once again from Eq.\eqref{ratef}, the full behavior of the rate function at two loops in the large field limit is given by:

\begin{align}\label{LFrate1}
  & I_{\zeta}(x)\sim_{x\rightarrow\infty} \frac{x^{4}}{u_{*}}\left(1+\epsilon\log{x}+\frac{\epsilon^{2}}{54}(25\log{x}+27\log^{2}x)\right)\nonumber \\
   &+\lim_{x\rightarrow\infty}\frac{1}{u_{*}}\left[\left(\frac{\epsilon}{6}+\frac{17}{162}\epsilon^{2}\right)\Delta(x^{2})+\frac{\epsilon^{2}}{12}x^{2}\log{x^{2}}\theta^{(1)}(x^{2})\right].
\end{align}
It has already been shown in \cite{sahu2024generalization, eisenriegler1987helmholtz}, that in the limit $x\rightarrow\infty$, the function $\Delta(x^{2})$ behaves as $\lim_{x\rightarrow\infty}\Delta(x^{2})\sim -\log{x^{2}}$. Using the techniques shown in the appendix of \cite{sahu2024generalization}, one can find an alternative representation of $\theta^{(1)}(x)$ given by:
\begin{align}\label{theta1A}
    \theta^{(1)}(x) & =\int_{1}^{\infty}d\sigma\left(\vartheta^{4}(\sigma)-1\right)e^{-x/\sigma}\nonumber\\
    &+\int_{1}^{\infty}d\sigma\left(\vartheta^{4}(\sigma)-1\right)e^{-x\sigma} -\frac{(1-e^{-x})}{x}-E_{2}(x),
\end{align}
with $E_{2}(z)=\int_{1}^{\infty}\frac{e^{-zt}}{t^{2}}$. Using Eq.\eqref{theta1A}, one can now show $\lim_{x\rightarrow\infty}x^{2}\theta^{(1)}(x^{2})\log{x^{2}}\sim-\log{x^{2}}$. Thus Eq.\eqref{LFrate1} now becomes:
\begin{align}\label{LFrate2}
    &I_{\zeta}(x)  \sim_{x\rightarrow\infty} \frac{x^{4}}{u_{*}}\left(1+\epsilon\log{x}+\frac{\epsilon^{2}}{54}(25\log{x}+27\log^{2}x)\right)\nonumber \\
   &-\left(1+\frac{\epsilon}{2}+O(\epsilon^{2})\right)\log{x}.
\end{align}
The subleading correction in Eq.\eqref{LFrate2} which goes as $-\left(1+\frac{\epsilon}{2}\right)\log{x}$ is nothing but $-\frac{\delta-1}{2}\log{x}$ expanded to first order in $\epsilon$. This suggests that the large field behaviour of the PDF or equivalently the rate function is universal and is entirely determined by one critical exponent $\delta$ with the analytical structure being $\lim_{x\rightarrow\infty}I_{\zeta}(x)\sim x^{\delta+1}-\frac{(\delta-1)}{2}\log{x}$. Equivalently, the PDF at the large field is given by $x^{\frac{(\delta-1)}{2}}e^{-x^{\delta+1}}$. This behaviour has already been seen at one-loop and has also been predicted for the Ising model some time ago in \cite{Hilfer1995, bruce1995critical}  and is in fact generic for equilibrium systems as shown in \cite{Stella2023, balog2024universal}.

\section{Comparison with Monte-Carlo Simulations}
For the comparison with the Monte-Carlo simulations we need to a priori get rid of some non-universal factors. Fixing these non-universal factors allows us to exactly compare the rate functions obtained from MC simulations and the ones computed from perturbation theory.\\

In \cite{bervillier1976universal, privman1984universal}, it has been shown that only two such amplitudes survive in the renormalized theory. These two scales are the temperature scale and the scale of the order parameter. Thus, after fixing these two scales, we should be able to compare the rate functions computed from perturbation theory and those computed from MC simulations for all values of $\zeta$. We fix these scales in the following way: the scale of the order parameter is fixed by demanding that the minimum of the rate function for $\zeta=0$ computed either from perturbation theory or from the MC data occurs at the same point. In our case, we have normalized so as to have both of them at $x=1$. For comparison at $\zeta=0$, the temperature scale need not be fixed. However, for comparison of rate functions for $\zeta\neq 0$, we must fix the temperature scale as well. 

We have already noticed in the one-loop calculation that there happens to be a value of $\zeta$, that we call $\zeta_{c}$, at which the rate function and consequently the PDF changes concavity at the origin and thus goes from simply peaked to doubly peaked. Mathematically speaking, this number $\zeta_{c}$ is given by the value of $\zeta$ for which the quantity $\frac{\partial I_{\zeta}(\rho)}{\partial \rho}|_{\rho=0}$ (with $\rho=\tilde{s}^{2}/2$) is vanishing. One of our motivations for studying these PDFs is also to predict this number $\zeta_{c}$ which is universal once the temperature scale is set. Thus the temperature scale is normalized such that the quantity $\frac{\partial I_{\zeta}(\rho)}{\partial \rho}|_{\rho=0}$ computed either from perturbation theory or from MC simulations at a normalization point $\zeta=\zeta_{N}$ is the same. In our case, we choose this to be at $\zeta_{N}=5$. Setting the temperature scale in the above mentioned way, we predict the value of $\zeta_{c} = 2.1$ which is in very good agreement with the predicted $2.05-2.1$ from the MC simulations. To check the robustness of our normalization procedure, we have also fixed the temperature scale at a different normalization point, $\zeta_{N}=3$. Doing this, we again find $\zeta_{c}$ to be around $2.1$, thereby suggesting that our normalization procedure is indeed a robust one. \\

\begin{figure}[t]
\centering
\includegraphics[width=\linewidth]{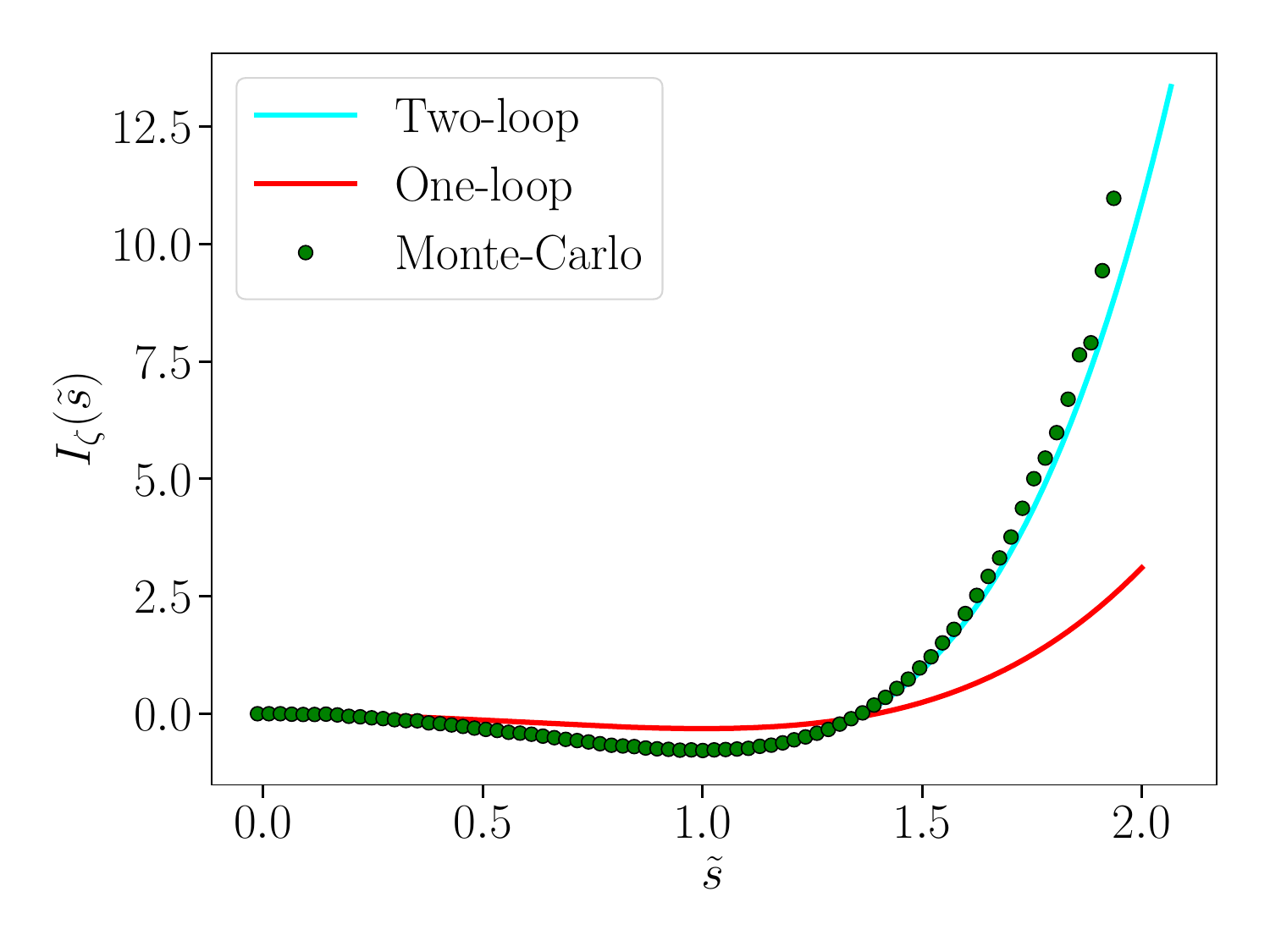}
\caption{ \label{Rate-zeta=02L.pdf} Comparison of $I_{\zeta=0}(x)$ obtained from Monte Carlo (MC) simulations (green) and from first (red) and second order (cyan) of perturbation theory. }
\end{figure}

\begin{figure}[t]
\centering
\includegraphics[width=\linewidth]{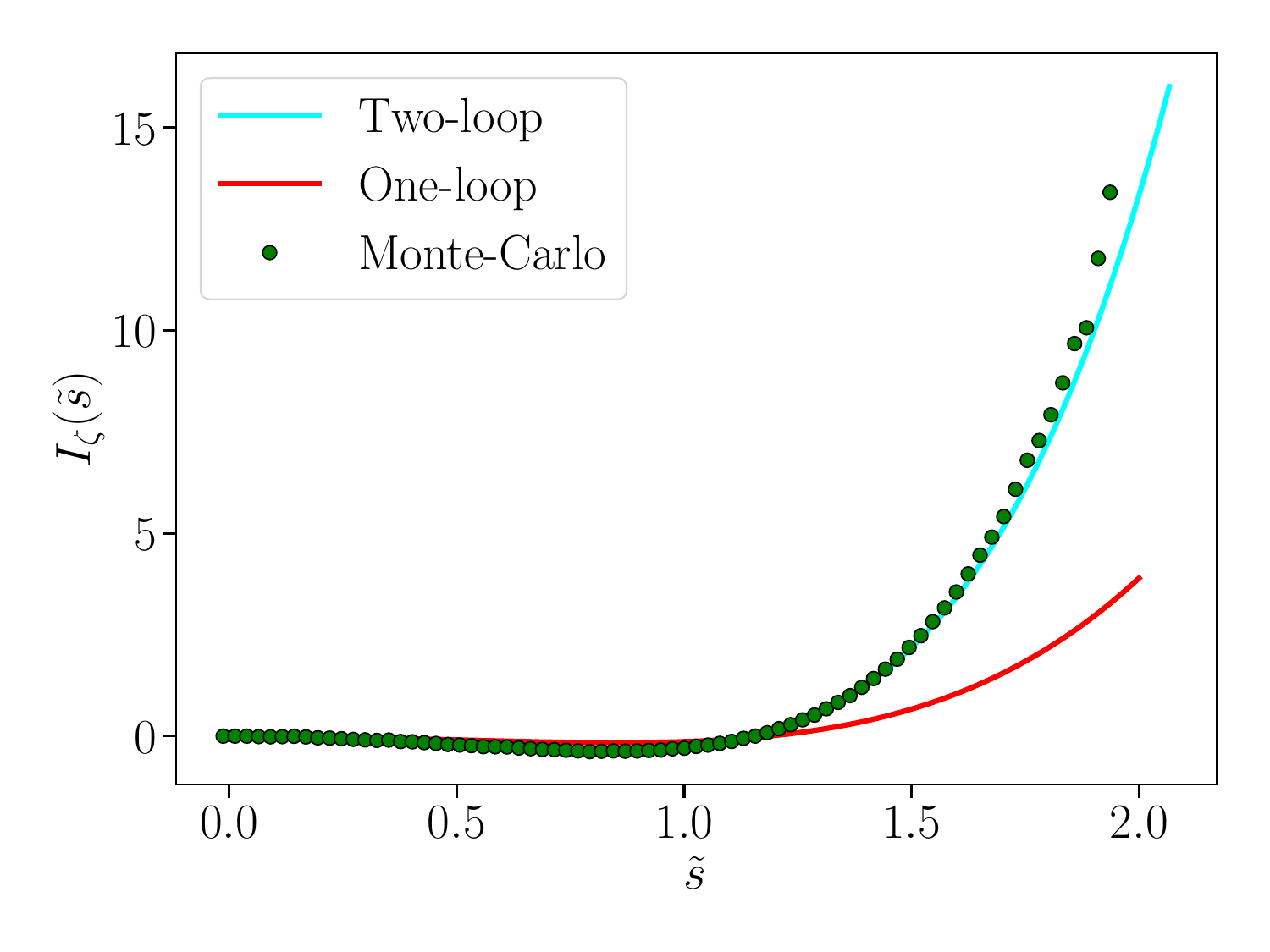}
\caption{\label{zeta=1}
Same as Fig.~\ref{Rate-zeta=02L.pdf}  with $\zeta=1$. }
\end{figure}

\begin{figure}[t]
\centering
\includegraphics[width=\linewidth]{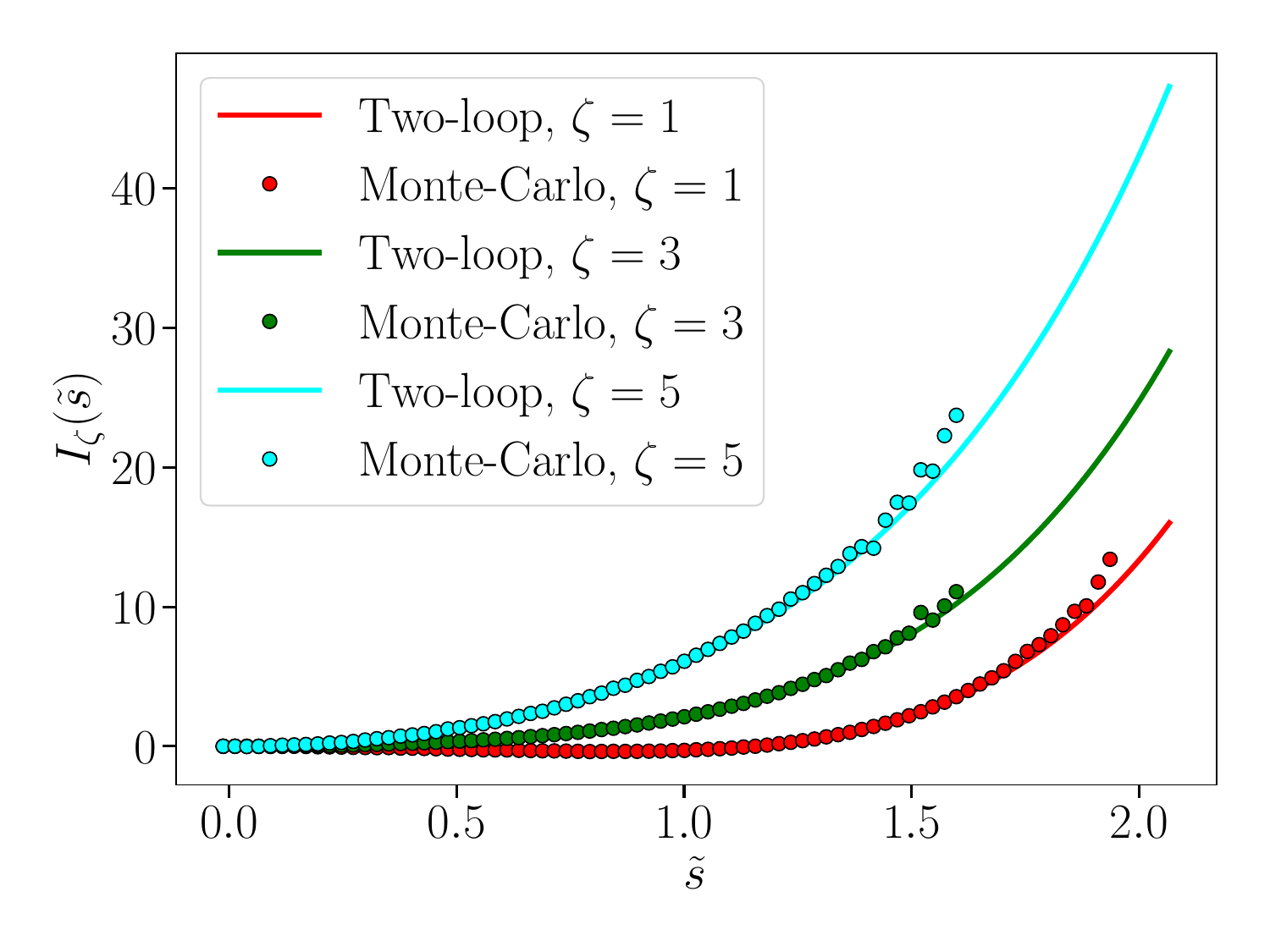}
\caption{\label{zeta 135} Comparison between the rate function $I_{\zeta}(\tilde{s})$ obtained at two-loops and from Monte-Carlo Simulations for $\zeta=1$ (red), $\zeta=3$ (green) and $\zeta=5$ (cyan).}
\end{figure}

\begin{figure}[t]
\centering
\includegraphics[width=\linewidth]{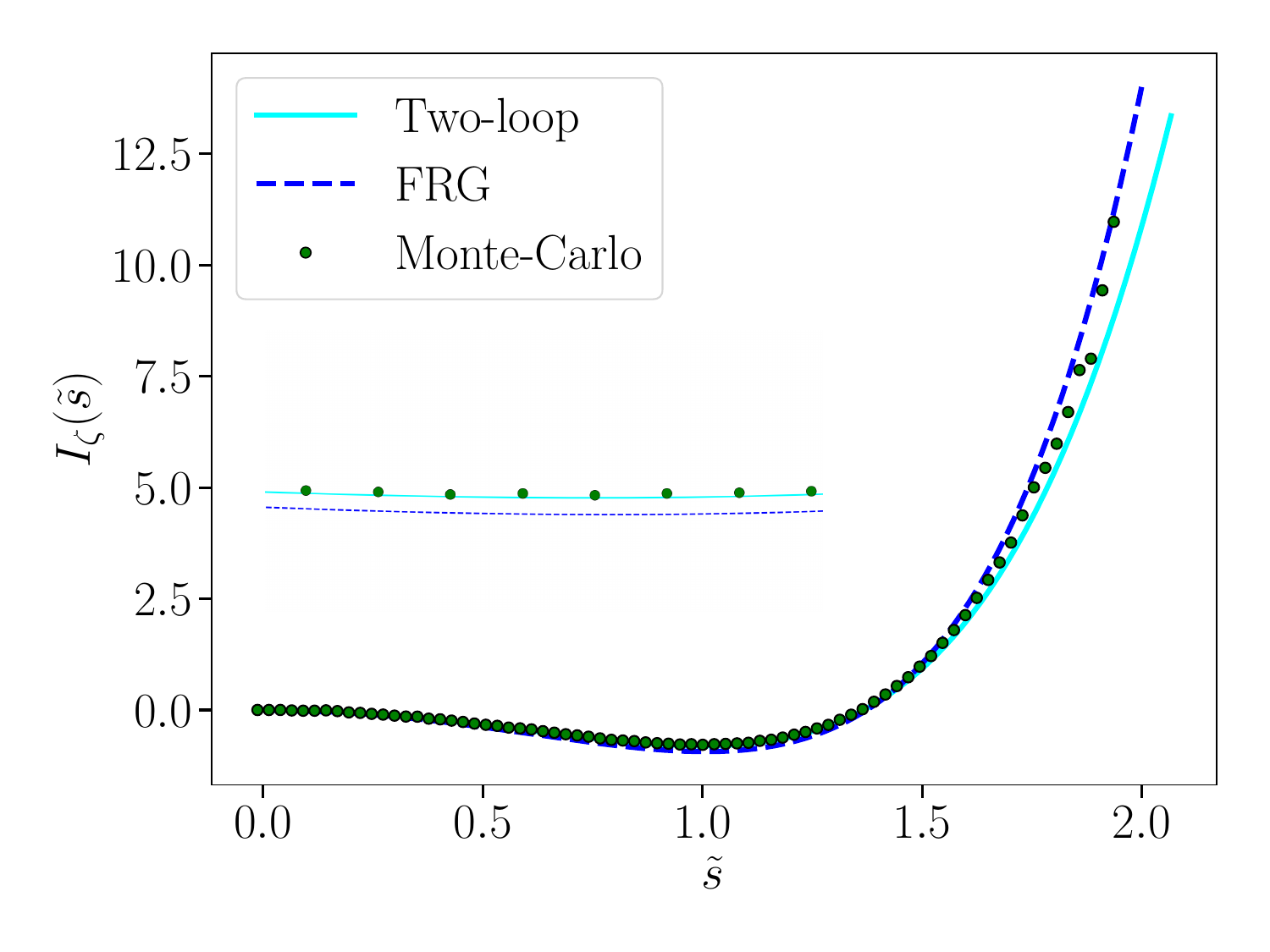}
\caption{ \label{Rate-zeta=02FL.pdf} Comparison of $I_{\zeta=0}(x)$ obtained from Monte Carlo (MC) simulations (green), second order of perturbation theory (cyan) and the FRG calculation (LPA approximation) from \cite{balog2022critical}. }
\end{figure}

What we find is that indeed after fixing these two scales, the agreement between the Monte-Carlo data and the two-loop results becomes remarkable as shown in Figs. \ref{Rate-zeta=02L.pdf}, \ref{zeta=1} and \ref{zeta 135}. We also stress the fact that at one loop, the prediction of the shape of the rate function is quite accurate which means that after one rescaling in the amplitude of the rate function the agreement between the MC simulations and the one-loop becomes rather good. Of course, this rescaling is forbidden as per the two-scale factor universality discussed in \cite{bervillier1976universal, privman1984universal}. In essence, at two loops what we also improve is the determination of the minimum of the rate function at $\zeta=0$. The determination of the minima of $I_{\zeta=0}(x)$ from different methods is shown in table \ref{tabImin}. We also compare our results with the existing results from the FRG calculations as shown in  Fig. \ref{Rate-zeta=02FL.pdf}. 

\subsection*{An aside: The two scale factor universality}

As stated before, only two scales are needed to be fixed as a consequence of the ``two-scale factor universality"  proven by Privman and Fisher \cite{privman1984universal}. As a consequence, using both $\zeta$ and the  scaling variable $\tilde s$  is sufficient to make $I_\zeta$, defined by $I_\zeta(\tilde s)=L^{d} I(s, \xi_{\infty}, L)$, universal. However, apriori one might assume that there are three scales that needs to be fixed for a proper comparison with the Monte-Carlo data i.e. the scale of $s$, $L$ and $\xi_{\infty}$ or temperature $t$ should be fixed separately and independently. In the following, we show how this assumption breaks down even in the the simple case of independent random variables. \\
As a toy example, consider $L^{d}$ independent random variables. One might think that building the universal rate function out of the scaling variable would require an extra normalization condition as the rate function appears together with a non-universal $L^d$ factor: 
\begin{equation}
    P(\hat{s}=s)\propto e^{-L^{d}I(s)}.
\end{equation}
with $I(s)=s^2/(2\sigma^2)$. However this is not true because going to the right scaling variable 
\begin{equation}
    \tilde{s}=L^{d/2}s/s_0
\end{equation}
is sufficient (i) to remove the non-universal $L^{d}$ pre-factor in front of $I(s)$ and (ii) to remove the non-universal $\sigma^2$ through the freedom of adjusting the normalization of  $\tilde s$, i.e. $s_0$. No other non-universal normalization condition is needed to get
\begin{equation}
    P(\hat{s}=\frac{1}{L^{d/2}}\tilde{s})\propto e^{-L^{d}I(s)}\propto e^{-{\tilde I}(\tilde{s})}
\end{equation}
where ${\tilde I}(\tilde{s})=\tilde s^2/2$ is universal. In this rather trivial example, only one normalization condition is needed and not two.

Of course, in the strongly correlated case, the situation is more complicated because even after extracting the factor $L^{d}$ as in Eq.~\eqref{pdf34} above, the function $I$ remains $\xi_\infty$ and $L$-dependent. However, as in the case of independent variables, once  $s$  has been traded for the scaling variable $\tilde{s}=sL^{(d-2+\eta)/2}/s_0$, the $L^{d}$ factor disappears leaving only the ratio $\zeta=\xi_{\infty}/L$ as the  only other independent variable in the simultaneous limit of infinite volume and criticality:
\begin{equation}
    L^{d}I(s, \xi_{\infty}, L) = I_\zeta(\tilde{s}).
\end{equation}
Once $s_0$  as well as the scale relating $\xi_{\infty}$ and $t\propto T-T_c$  are fixed, no extra normalization condition is required to make $I_\zeta$ universal and thus only two such conditions are required and not three.\\

\begin{table}[t]
  \begin{center}
    \begin{tabular}{ c|c } 
      \text{Methods} 
       &$I^{min}_{\zeta=0}(x)$\\
      \hline
       \\
 \text{  Mean Field} & $0$ \\
     &\\
       \text{One-loop} \cite{sahu2024generalization} & $-0.317$\\
     &\\
     \text{Two-loop} & $-0.805$\\
     &\\
      \text{Monte-Carlo} & $-0.783$\\
     &\\
       \text{FRG} (LPA) \cite{balog2022critical} & $-0.979$\\
        &\\
       \text{FRG} (DE2) \cite{Private} & $-0.730$\\
    \end{tabular}
    \caption{\label{tabImin} Determination of the minimum of the rate function at $I^{min}_{\zeta=0}(x)$ from different methods.}
  \end{center}
\end{table}

\section{Conclusions}
In this paper, we present a two-loop perturbative calculation of the rate-function and consequently the PDF of the sum of the properly normalized spins (i.e. the PDF of the magnetization mode) of the Ising model in $d=4-\epsilon$. Just as for the one-loop case, we find that in the simultaneous limit of criticality $\xi_{\infty}\rightarrow \infty$ and infinite volume $L\rightarrow\infty$, there exists not one but infinitely many PDFs parameterized by $\zeta$ with $\zeta= \lim_{L\rightarrow\infty}\frac{L}{\xi_{\infty}}$. While at one-loop we correctly predicted the shape of the rate function, we were far from any kind of quantitative accuracy. At two loops, we find that this improves massively, and the two-loop results agree very well with the Monte-Carlo simulations.\\

We see that in the two-loop case, the agreement with the Monte-Carlo results in the tails is also much better compared to the one-loop, the reason being that the tail of the rate function is largely determined by one exponent, i.e. $\delta$, which is also determined with a better accuracy at two loops. However, it is still possible to improve the two-loop result at large fields using RG as shown in \cite{guida19973d, chung1999renormalization}. This is beyond the scope of the current article but we do wish to cure the large field behaviour using RG in our future works. We also notice, that unlike the one-loop case, the generalization of the two-loop results to the $O(N)$ vector model is not as simple, since it requires computation of more diagrams, which we also plan to address in our upcoming works. We also found that at one-loop the generalization of our result to the low temperature phase i.e. $\zeta<0$ was rather trivial, whereas this does not seem to be the case for the two-loop computation and hence would require further investigations.\\
Another question that remains unanswered is the shape of these PDFs exactly in $d=4$. We already discussed how the current method does not directly allow us to study these PDFs at $d=4$ and hence  developing a method that does, also peeks our interest. A natural question that arises out of our work is if there exists such a generalized limiting law for the stochastic variables in out of equilibrium systems like directed percolation, reaction-diffusion systems etc.   
 All in all, the current study provides a gateway to study a host of different problems in the context of probability distribution functions of strongly correlated variables within a perturbative setting.

\section{Acknowledgements}
The author would like to thank Bertrand Delamotte for his support during the work and for his valuable comments on the initial draft of this manuscript. The author would also like to thank Adam Ran\c con for his comments on the manuscript. The author also thanks Adam Ran\c con and Ivan Balog for many fruitful discussions and for providing him with the Monte-Carlo data.

\newpage
\appendix
\onecolumngrid
\section{}\label{appenA}
In the calculation that follows, we show how the rate function $I({s})$ defined as $P(\hat{s}=s)=Ne^{-L^dI({s})}$, is associated with $\lim_{M\rightarrow\infty}\Gamma_M(\phi(x)=s)$.
\begin{align}
    & e^{-\Gamma_{M}[\phi(x)]} =\mathcal{N}\int  D\hat{\phi}~e^{-\int \mathcal{H}[\hat{\phi}]+\int \frac{\delta \Gamma_{M}}{\delta\phi}.({\hat\phi}-\phi)}\exp(-\frac{M^2}{2}\left(\int_{x}(\hat{\phi}(x)-\phi)\right)^2)\nonumber\\
    & \implies e^{-\Gamma_{M}[\phi(x)=s]} = \mathcal{N}\int D\hat{\phi}~e^{-\int \mathcal{H}[\hat{\phi}]+h.\int (\hat{\phi}-s)}\exp(-\frac{M^2}{2}\left(\int_{x}({\hat\phi}(x)-s)\right)^2)\nonumber\\
     & \implies \lim_{M\rightarrow\infty} e^{-\Gamma_{M}[\phi(x)=s]} = \mathcal{N}\int D\hat{\phi}~e^{-\int \mathcal{H}[\hat{\phi}]+h.\int (\hat{\phi}-s)}\delta\left(\int{\hat\phi(x)}-s \right)\nonumber\\
     & \implies \lim_{M\rightarrow\infty} e^{-\Gamma_{M}[\phi(x)=s]} = \mathcal{N}\int D\hat{\phi}~\delta\left(\int{\hat\phi(x)}-s \right)e^{-\int \mathcal{H}[\hat{\phi}]}\nonumber
\end{align}
Thus we see that: $P(\hat{s}=s)\propto\lim_{M\rightarrow\infty} e^{-\Gamma_{M}[\phi(x)=s]}$, meaning that $\Gamma_{M}[\phi(x)=s]=L^dI({s})$.
\section{}\label{appenB}
In this section, we show how to explicitly compute the modified effective potential at two loops diagrammatically. We start off by recalling the one-loop calculation of $Z_{M, s}[h]$ and thereby subsequently computing $\Gamma_{M}[\phi]$ at one loop and then at two loops. $Z_{M, s}[h]$ at one loop is given by : 
 \begin{align}\label{b}
     \mathcal{Z}_{M, s}[h] &=\mathcal{N}\int D\hat{\phi}\exp(- \mathcal{H}_{M}[\hat{\phi}]+\int h.\hat{\phi})\nonumber\\
     & =  \mathcal{N}\int D\hat{\phi}'\exp(- \mathcal{H}_{M, s}[\hat{\phi}_{0}+\hat{\phi}']+\int h(\hat{\phi}_{0}+\hat{\phi}'))\nonumber\\
     & = \mathcal{N}\int D\hat{\phi}'\exp(- \left(\mathcal{H}_{M, s}[\hat{\phi}_{0}]-\int h~\hat{\phi}_{0}\right)-\int\left(\frac{\delta \mathcal{H}_{M, s}}{\delta\hat{\phi}}|_{\hat{\phi}=\hat{\phi}_{0}} -h\right)\hat{\phi}'-\frac{1}{2}\int_{x_{1}, x_{2}}\mathcal{H}^{(2)}_{M, s}(x_{1},x_{2})\hat{\phi}'(x_{1})\hat{\phi}'(x_{2})+......)\nonumber\\
      & = \mathcal{N}\exp(- \mathcal{H}_{M, s}[\hat{\phi}_{0}]+\int h~\hat{\phi}_{0}-\frac{1}{2}\Tr\log\mathcal{H}^{(2)}_{M, s}).
\end{align}
 From Eq.\eqref{phi} at one-loop one ends up with:
  \begin{equation}\label{c}
      \phi(x)=\hat{\phi}_{0}(x) .
  \end{equation}
  Using Eqs.\eqref{PE1}, (\ref{b}) and (\ref{c}), one thus derives:
  \begin{equation}\label{1LPE}
      \Gamma_{M}[\phi] =\mathcal{H}[\phi]+\frac{1}{2}\Tr\log\mathcal{H}^{(2)}_{M, s}[\phi]-\log[\mathcal{N}].
  \end{equation}
  Here $\log[\mathcal{N}]$ has has been chosen such that it is equal to the value of $\mathcal{H}[\phi]+\Tr\log\mathcal{H}^{(2)}_{M}[\phi]$ computed at vanishing field and vanishing mass (i.e. at $T=T_{c}$) with an addition of a constant-infinite pole counter-term to cancel out the divergence coming out from the one-loop.
  
We start with the two-loop computation of the quantity $W_{M,s}[h]$. In the derivation that proceeds, we use an auxiliary $\hbar$ as means of counting loops. The full $W_{M,s}[h]$ up to all orders of perturbation theory is given by:
\begin{align}\label{PE2}
W_{M,s}[h] &=-\mathcal{H}_{M,s}[\hat{\phi}_{0}]+ \int_{x} h(x)\hat{\phi}_{0}(x)-\frac{\hbar}{2}\Tr\log\mathcal{H}^{(2)}_{M, s}[\hat{\phi}_{0}]
 +W^{(2)}_{M,s}[h]-\log{[\mathcal{N}]},
\end{align}
where $W^{(2)}_{M,s}[h]$ is given by:
\begin{align}
W^{(2)}_{M,s}[h]=\hbar\log{Z^{(2)}_{M,s}[h]},
\end{align}
with 
\begin{align}\label{Z2}
Z^{(2)}_{M,s}[h]=\frac{\int \mathcal{D\hat{\phi}}\exp{-\frac{1}{\hbar}\left(\mathcal{H}_{M,s}[\hat{\phi}+\hat{\phi}_{0}]-\int_{x} h(x))\hat{\phi}_{0}(x)-\mathcal{H}_{M,s}[\hat{\phi}]\right)}}{\int \mathcal{D\hat{\phi}}\exp{-\frac{1}{\hbar}\left(\int\hat{\phi}.\frac{\delta^2\mathcal{H}_{M,s}}{\delta\hat{\phi}^2}|_{\hat{\phi}=\hat{\phi}_{0}}.\hat{\phi}\right)}}
\end{align}
[Here $\log{\mathcal{N}}$ is now modified so as to include constant pole terms of higher orders as well]. Keep in mind \eqref{PE2} is true up to all orders in perturbation theory. If we just want to consider terms up to two-loop order, one could expand \eqref{Z2} up to the fourth order in $\hat{\phi}$. Thus for the computation of the effective potential to two-loop order we shall just consider:
\begin{align}\label{Z2-2loop}
Z^{(2)}_{M,s}[h]=\left\langle\exp{-\frac{1}{\hbar}\left(\frac{1}{4!}\int_{x, y, z, w}\hat{\phi}(x)\hat{\phi}(y)\hat{\phi}(z)\hat{\phi}(w)\frac{\delta^4\mathcal{H}_{M,s}}{\delta\hat{\phi}^4}|_{\hat{\phi}=\hat{\phi}_{0}}+\frac{1}{3!}\int_{x, y, z}\hat{\phi}(x)\hat{\phi}(y)\hat{\phi}(z)\frac{\delta^3\mathcal{H}_{M,s}}{\delta\hat{\phi}^3}|_{\hat{\phi}=\hat{\phi}_{0}}\right)}\right\rangle_{L},
\end{align}
where:-
\begin{align}\label{Z2k-2loop}
\left\langle K \right\rangle_{L}=\frac{\int \mathcal{D\hat{\phi}}\exp{-\frac{1}{\hbar}\left(\int_{x, y}\hat{\phi}(x).\frac{\delta^2\mathcal{H}_{M,s}}{\delta\hat{\phi}^2}|_{\hat{\phi}=\hat{\phi}_{0}}.\hat{\phi}(y)+K\right)}}{\int \mathcal{D\hat{\phi}}\exp{-\frac{1}{\hbar}\left(\int_{x, y}\hat{\phi}(x).\frac{\delta^2\mathcal{H}_{M,s}}{\delta\hat{\phi}^2}|_{\hat{\phi}=\hat{\phi}_{0}}.\hat{\phi}(y)\right)}}.
\end{align}
Next, we also introduce the quantity:
\begin{equation}\label{Gamma1}
\Gamma^{(1)}_{ M,s}[h]=\frac{\hbar}{2}\Tr\log{\mathcal{H}^{(2)}_{M,s}}-W^{(2)}_{M,s}[h].
\end{equation} 

Thus from \eqref{PE2} we have:
\begin{align}\label{PE3}
\frac{\delta W_{M,s}[h]}{\delta h} & = \left[h-\frac{\delta \mathcal{H}_{M,s}[\hat{\phi}]}{\delta \hat{\phi}}|_{\hat{\phi}=\hat{\phi}_{0}}\right]\frac{\delta \hat{\phi}_{0}}{\delta h}+\hat{\phi}_{0}-\frac{\delta \Gamma^{(1)}_{ M,s}[h]}{\delta h}\nonumber\\
 \implies  \hat{\phi}_{0} & =\phi+\phi^{(1)},
\end{align}
 with:
 \begin{equation}\label{phi1}
 {\phi}^{(1)} = \frac{\delta \Gamma^{(1)}_{ M,s}[h]}{\delta h}.
 \end{equation}
 Where in the last line of \eqref{PE3}, we have made use of the relation Eq.\eqref{phi}. One must note that $\phi^{(1)}$ is itself of the order of $\hbar$. Also note that $\phi^{(1)}$ depends on $\hat{\phi}_{0}$ through $h$. Since for two-loop contribution of the 'modified effective potential' we only consider $\hat{\phi}_{0}$ to one-loop order, $\phi^{(1)}$ can be considered just as a functional of $\phi$. 
 
Plugging back \eqref{PE3} in Eq.\eqref{PE1} one obtains:
 
\begin{align}\label{2LP}
 \Gamma_{M}[\phi]+W_{M,s}[h] & =\int h(\phi+\phi^{(1)})~\phi-\frac{M^2}{2}\left(\int_{x}\phi-s\right)^2\nonumber\\
\implies \Gamma_{M}[\phi] & = \int h[\phi+\phi^{(1)}]~\phi-\frac{M^2}{2}\left(\int_{x}\phi-s\right)^2+\mathcal{H}_{M, s}[\hat{\phi}_{0}]-\int h.\phi_{0}+\Gamma^{(1)}_{M, s}[h]-\log[\mathcal{N}]\nonumber\\
& =\mathcal{H}_{M, s}[\phi+\phi^{1}] -\int\phi^{(1)}h[\phi+\phi^{(1)}]+\Gamma^{(1)}_{M, s}[\phi+\phi^{(1)}]-\frac{M^2}{2}\left(\int_{x}\phi-s\right)^2-\log[\mathcal{N}]\nonumber\\
& = \mathcal{H}[\phi]+\int \phi^{(1)}\left(\frac{\delta \mathcal{H}_{M,s}}{\delta \phi}-h[\phi]\right)+\int\phi^{(1)}\frac{\delta^{2}\mathcal{H}_{M,s}}{\delta \phi^2}\phi^{(1)}-\int \phi^{(1)}\frac{\delta h}{\delta \phi} \phi^{(1)}\nonumber\\
& +\Gamma^{(1)}_{M, s}[\phi]+\int \phi^{(1)}\frac{\delta \Gamma^{(1)}_{M, s}}{\delta \phi}-\log[\mathcal{N}]+O(\hbar^3),
\end{align} 
Now we focus on computing the second term of  Eq. \eqref{2LP} upto order $O(\hbar^3)$. For the calculations ahead we define the inverse propagator $\mathcal{D}^{-1}$ as follows:
\begin{equation}\label{D2}
\frac{\delta^{2}\mathcal{H}_{M,s}}{\delta \phi^2}=\mathcal{D}^{-1}.
\end{equation}
 Using \eqref{1LPE} and derivating it w.r.t. $\phi$ and expanding to order $\hbar$, one finds:
 \begin{align}\label{IC}
 & \frac{\delta \Gamma_{M}[\phi]}{\delta \phi}=\frac{\delta \mathcal{H}}{\delta \phi}+\frac{\hbar}{2}\Tr\left(\mathcal{D}\frac{\delta }{\delta \phi}\mathcal{D}^{-1}\right)\nonumber\\
 & \implies h(\hat{\phi}_{0})-M^2\left(\int_{x}\phi -s\right) = \frac{\delta  \mathcal{H}}{\delta \phi}+\frac{\hbar}{2}\Tr(\mathcal{D}\frac{\delta}{\delta\phi}\mathcal{D}^{-1})\nonumber\\
 & \implies h[\phi]+\int \phi^{(1)}\frac{\delta h}{\delta \phi}-M^2\left(\int_{x}\phi -s\right)=\frac{\delta \mathcal{H}}{\delta \phi}+\frac{\hbar}{2}\Tr\left(\mathcal{D}\frac{\delta }{\delta \phi}\mathcal{D}^{-1}\right)\nonumber\\
 & \implies \frac{\delta \mathcal{H}_{M,s}}{\delta \phi}-h[\phi]= \phi^{(1)}.\frac{\delta h}{\delta\phi}-\frac{\hbar}{2}\Tr\left(\mathcal{D}\frac{\delta}{\delta\phi}\mathcal{D}^{-1}\right).
 \end{align}
 Where in the last line we have used the fact: $\mathcal{H}_{M, s}[\hat{\phi}]=\mathcal{H}[\hat{\phi}]+ \frac{M^2}{2}\left(\int_{x}(\hat{\phi}(x)-s)\right)^2$. 
 Using \eqref{phi1}, we also get:
 \begin{equation}\label{phi2}
 \phi^{(1)} =\frac{\delta \Gamma^{(1)}_{M, s}}{\delta \hat{\phi}_{0}}.\frac{\delta\hat{\phi}_{0}}{\delta h}.
\end{equation}  
 Using \eqref{Gamma1}, we also have:
 \begin{equation}\label{Ms}
 \frac{\delta \Gamma^{(1)}_{M, s}}{\delta \hat{\phi}_{0}}=\frac{\hbar}{2}\Tr\left(\mathcal{D}\frac{\delta}{\delta \hat{\phi}_{0}}\mathcal{D}^{-1}\right).
 \end{equation}
We also know that:
\begin{align}
\frac{\delta \mathcal{H}_{M, s}}{\delta \hat{\phi}}|_{\hat{\phi}=\hat{\phi}_{0}}=h,
\end{align}
from which it is easy to deduce:
\begin{equation}\label{Ms1}
\frac{\delta^{2} \mathcal{H}_{M, s}}{\delta \hat{\phi}^2}|_{\hat{\phi}=\hat{\phi}_{0}}=\frac{\delta h}{\delta\hat{\phi}_{0}}.
\end{equation}
 
 As mentioned earlier, for the two-loop computation of the effective potential, it suffices to compute $\phi^{(1)}$ to the order $\hbar$. Thus one can easily trade in $\hat{\phi}_{0}$ for $\phi$ in \eqref{Ms} and \eqref{Ms1} and then using \eqref{phi2} one obtains:
 
 \begin{equation}\label{hbarphi}
 \phi^{(1)}=\frac{\hbar}{2}\Tr\left(\mathcal{D}\frac{\delta}{\delta \phi}\mathcal{D}^{-1}\right).\mathcal{D}+O(\hbar  ^2).
 \end{equation}
 Trading in $\phi_{0}$ for $\phi$ in \eqref{Ms1}, one also obtains:
 \begin{equation}
 \frac{\delta h}{\delta \phi} = \mathcal{D}^{-1}+O(\hbar).
 \end{equation}
 Thus we have :
 \begin{equation}\label{IC1}
 \phi^{(1)}.\frac{\delta h}{\delta\phi}= \frac{\hbar}{2}\Tr\left(\mathcal{D}\frac{\delta}{\delta \phi}\mathcal{D}^{-1}\right)+O(\hbar  ^2).
 \end{equation}
 
 Using \eqref{IC} and \eqref{IC1} one thus obtains:
 \begin{equation}\label{h^3}
 \frac{\delta \mathcal{H}_{M,s}}{\delta \phi}-h[\phi]= O(\hbar^2).
 \end{equation}
Thus the second term in \eqref{2LP} is of the order $O(\hbar^3)$ and hence doesn't contribute to the two loop calculation of the effective potential. Now we have all the ingridients at our disposal to compute $\Gamma_{M}[\phi]$ in two loops. We start by computing \eqref{2LP} up to order $O(\hbar ^2)$ in its full glory. Using \eqref{h^3} in \eqref{2LP}, we find:
\begin{align}\label{2LP1}
\Gamma_{M}[\phi] & = \mathcal{H}[\phi]+\int\phi^{(1)}\frac{\delta^{2}\mathcal{H}_{M,s}}{\delta \phi^2}\phi^{(1)}-\int \phi^{(1)}\frac{\delta h}{\delta \phi} \phi^{(1)}\nonumber\\
& +\Gamma^{(1)}_{M, s}[\phi]+\int \phi^{(1)}\frac{\delta \Gamma^{(1)}_{M, s}}{\delta \phi}-\log[\mathcal{N}]+O(\hbar^3).
\end{align}

Now using \eqref{IC1}, \eqref{Gamma1} and \eqref{hbarphi} in \eqref{2LP1}, we get:

\begin{align}\label{GammaP}
\Gamma_{M}[\phi] & =  \mathcal{H}[\phi]+\frac{\hbar}{2}\Tr\log{\mathcal{H}^{(2)}_{M,s}}-W^{(2)}_{M,s}[h]+\frac{\hbar^2}{8}\Tr\left(\mathcal{D}\frac{\delta}{\delta \phi}\mathcal{D}^{-1}\right).\mathcal{D}.\Tr\left(\mathcal{D}\frac{\delta}{\delta \phi}\mathcal{D}^{-1}\right)-\log[\mathcal{N}].
\end{align}

The next step is to compute $W^{(2)}_{M, s}[h]$ upto two-loop order. Since we know  $W^{(2)}_{M, s}[h]$ is the generator of connected correlation functions, in essence the job at our hand reduces to computing the connected diagrams arising out of   \eqref{Z2-2loop}. This entails in computing the following:
\begin{align}\label{wick}
W^{(2)}_{M,s}[h] & =\hbar\left\langle\exp{-\frac{1}{\hbar}\left(\frac{1}{4!}\int_{x, y, z, w}\hat{\phi}(x)\hat{\phi}(y)\hat{\phi}(z)\hat{\phi}(w)\frac{\delta^4\mathcal{H}_{M,s}}{\delta\hat{\phi}^4}|_{\hat{\phi}=\hat{\phi}_{0}}+\frac{1}{3!}\int_{x, y, z}\hat{\phi}(x)\hat{\phi}(y)\hat{\phi}(z)\frac{\delta^3\mathcal{H}_{M,s}}{\delta\hat{\phi}^3}|_{\hat{\phi}=\hat{\phi}_{0}}\right)}\right\rangle_{C}\nonumber\\
& = \hbar\left\langle -\exp{\left(\frac{\hbar}{4!}\int_{x, y, z, w}\hat{\phi}(x)\hat{\phi}(y)\hat{\phi}(z)\hat{\phi}(w)\frac{\delta^4\mathcal{H}_{M,s}}{\delta\hat{\phi}^4}|_{\hat{\phi}=\hat{\phi}_{0}}+\frac{\hbar^{1/2}}{3!}\int_{x, y, z}\hat{\phi}(x)\hat{\phi}(y)\hat{\phi}(z)\frac{\delta^3\mathcal{H}_{M,s}}{\delta\hat{\phi}^3}|_{\hat{\phi}=\hat{\phi}_{0}}\right)}\right\rangle_{C}\nonumber\\
& = \hbar\left\langle -\frac{\hbar}{4!}\int_{x, y, z, w}\hat{\phi}(x)\hat{\phi}(y)\hat{\phi}(z)\hat{\phi}(w)\frac{\delta^4\mathcal{H}_{M,s}}{\delta\hat{\phi}^4}|_{\hat{\phi}=\hat{\phi}_{0}}\right\rangle_{C}\nonumber\\ &+\frac{\hbar}{2}\left\langle\frac{\hbar}{3!3!}\int_{x_{1}, y_{1}, z_{1}, x_{2}, y_{2}, z_{2}}\hat{\phi}(x_{1})\hat{\phi}(y_{1})\hat{\phi}(z_{1})\frac{\delta^3\mathcal{H}_{M,s}}{\delta\hat{\phi}^3}|_{\hat{\phi}=\hat{\phi}_{0}}\hat{\phi}(x_{2})\hat{\phi}(y_{2})\hat{\phi}(z_{2})\frac{\delta^3\mathcal{H}_{M,s}}{\delta\hat{\phi}^3}|_{\hat{\phi}=\hat{\phi}_{0}}\right\rangle_{C}+O(\hbar^3).
\end{align}
Using Wick's theorem and \eqref{D2} in \eqref{wick}, one hence obtains:
\begin{align}
W^{(2)}_{M,s}[h] & =\frac{1}{2}\frac{\hbar^2}{3!3!}\left\langle\int_{x, y, z}\hat{\phi}(x_{1})\hat{\phi}(y_{1})\hat{\phi}(z_{1})\frac{\delta\mathcal{D}^{-1}}{\delta\hat{\phi}_{0}}\hat{\phi}(x_{2})\hat{\phi}(y_{2})\hat{\phi}(z_{2})\frac{\delta\mathcal{D}^{-1}}{\delta\hat{\phi}_{0}}\right\rangle_{C}\nonumber\\
&-\frac{\hbar^2}{4!}\left\langle \int_{x, y, z, w}\hat{\phi}(x)\hat{\phi}(y)\hat{\phi}(z)\hat{\phi}(w)\frac{\delta^2\mathcal{D}^{-1}}{\delta\hat{\phi_{0}}^2}|_{\hat{\phi}=\hat{\phi}_{0}}\right\rangle_{C}+O(\hbar^3)\nonumber\\
& = \frac{\hbar^2}{12}\mathcal{I}_{1}+\frac{\hbar^2}{8}\mathcal{I}_{2}-\frac{\hbar^2}{8}\mathcal{I}_{3},
\end{align}

with:

\begin{tikzpicture}
    \begin{feynman}
     \vertex(a);
     \vertex [right =of a] (b); 
     \vertex [left=0.3em of a] {$\mathcal{I}_{1}=
$}; 
\vertex [right=0.4em of b] {$\mathcal{I}_{2}=
$};
 \diagram{      
            (a) -- (b); 
            (a) --(b);
            (a) --[half left](b);
            (a) --[half right](b);
            
        };

    \end{feynman}
\end{tikzpicture} 
 \begin{tikzpicture}
    \begin{feynman}
     \vertex(a);
     \vertex [right =of a] (b); 
      \vertex [above =of a] (c);
      \vertex [above =of b] (d);
 \diagram{      

            (a) -- [quarter left](c);
            (a) -- [quarter right](c);
            (a) -- (b);
            (b) -- [quarter left](d);
            (b) -- [quarter right](d);

        };

    \end{feynman}
    \end{tikzpicture}
\begin{tikzpicture}

    \begin{feynman}
     \vertex(a);
     \vertex [right =of a] (b); 
      \vertex [right =of b] (c);
      \vertex [left=0.2em of a] {$\mathcal{I}_{3}=
$};
      
        \diagram{
            
            (a) [scalar]--[half right](b);
            (a) --[half left](b);
            (b) --[half left](c);
            (b) --[half right](c);
        };

    \end{feynman}
    
   \end{tikzpicture}
      
   It's clear that algebraically $\mathcal{I}_{2}$ can be written as:
   
   \begin{equation}
   \mathcal{I}_{2}= \Tr\left(\mathcal{D}\frac{\delta}{\delta \phi}\mathcal{D}^{-1}\right).\mathcal{D}.\Tr\left(\mathcal{D}\frac{\delta}{\delta \phi}\mathcal{D}^{-1}\right).
\end{equation}
    
(Here we have directly traded in $\hat{\phi}_{0}$ for $\phi$, since they differ by a one-loop term.)
 
 What we find is rather expected i.e. the connected diagrams which are not 1-PI's  get exactly cancelled out by the terms appearing as a result of the fact that $\phi$ at two-loops is not equal to the mean field $\hat{\phi}_{0}$.
 
 Thus \eqref{GammaP} becomes:
 
 \begin{equation}\label{Gammarate}
 \Gamma_{M}[\phi]  =  \mathcal{H}[\phi]+\frac{\hbar}{2}\Tr\log{\mathcal{H}^{(2)}_{M,s}}+\frac{\hbar^2}{8}\mathcal{I}_{3}-\frac{\hbar^2}{12}\mathcal{I}_{1}-\log[\mathcal{N}].
 \end{equation}

\section{}\label{appen C} 
In this section, we show how to compute the diagrams emerging at two loops. We start by computing the sunset like term:
\begin{align}\label{sun}
& \frac{1}{L^{2d}}\left(\sum_{\substack{\{\vec{p},\vec{q}\}\neq 0,\\ \vec{p}\neq -\vec{q}}}\frac{1}{(\vec{p}^2+m^2)(\vec{q}^2+m^2)((\vec{p}+\vec{q})^2+m^2)}\right)\nonumber\\
& =  \frac{1}{L^{2d}}\left(\sum_{\{\vec{p},\vec{q}\}\neq 0}\frac{1}{(\vec{p}^2+m^2)(\vec{q}^2+m^2)((\vec{p}+\vec{q})^2+m^2)}-\frac{1}{m^2}\sum_{\vec{p}\neq 0}\frac{1}{(\vec{p}^2+m^2)^2}\right)\nonumber\\
& =\frac{1}{L^{2d}}\left(\sum_{\vec{p},\vec{q}}\frac{1}{(\vec{p}^2+m^2)(\vec{q}^2+m^2)((\vec{p}+\vec{q})^2+m^2)}-\frac{3}{m^2}\sum_{\vec{p}\neq 0}\frac{1}{(\vec{p}^2+m^2)^2}-\frac{1}{(m^2)^3}\right).
\end{align}  
In \eqref{sun}, the second term is rather easy to compute, and we will show how to compute terms like these later in the section. Our principle aim would be to compute the first term and extract the divergences out of it to perform a sensible renormalization. The method we use follows from \cite{bijnens2014two}. Using Poisson summation technique on the first term of \eqref{sun}, one thus obtains:\\

\begin{align}\label{sun1}
& \frac{1}{L^{2d}}\sum_{\vec{p},\vec{q}}\frac{1}{(\vec{p}^2+m^2)(\vec{q}^2+m^2)((\vec{p}+\vec{q})^2+m^2)}\nonumber\\
& = \int \frac{d^d{p}}{(2\pi)^d}~\frac{d^{d}{q}}{(2\pi)^d}\sum_{l_{p},l_{q}}\frac{e^{i l_{p}.\vec{p}+i l_{q}.\vec{q}}}{(\vec{p}^2+m^2)(\vec{q}^2+m^2)((\vec{p}+\vec{q})^2+m^2)}\nonumber\\
& = \int \frac{d^d{p}}{(2\pi)^d}~\frac{d^{d}{q}}{(2\pi)^d}\frac{1}{(\vec{p}^2+m^2)(\vec{q}^2+m^2)((\vec{p}+\vec{q})^2+m^2)}\nonumber\\
& +2 \int \frac{d^d{p}}{(2\pi)^d}~\frac{d^{d}{q}}{(2\pi)^d}\sum_{l_{p}\neq 0}\frac{e^{i l_{p}.\vec{p}}}{(\vec{p}^2+m^2)(\vec{q}^2+m^2)((\vec{p}+\vec{q})^2+m^2)}\nonumber\\
& + \int \frac{d^d{p}}{(2\pi)^d}~\frac{d^{d}{q}}{(2\pi)^d} \sum_{l_{p}\neq 0}\frac{e^{i l_{p}.(\vec{p}+\vec{q})}}{(\vec{p}^2+m^2)(\vec{q}^2+m^2)((\vec{p}+\vec{q})^2+m^2)}\nonumber\\
& + \int \frac{d^d{p}}{(2\pi)^d}~\frac{d^{d}{q}}{(2\pi)^d}\sum_{\substack{\{{l_{p}},{l_{q}}\}\neq 0,\\ {l_{p}}\neq {l_{q}}}}\frac{e^{i l_{p}.\vec{p}+i l_{q}.\vec{q}}}{(\vec{p}^2+m^2)(\vec{q}^2+m^2),((\vec{p}+\vec{q})^2+m^2)}
\end{align}

where $\{l_{p},l_{q}\}\in \{n L, m L\}$ with ${n, m}\in \mathbb{Z}^{d}$. Thus, we have separated the part coming from the continuum (first term in \eqref{sun1}) and the part encoding the finite size corrections (other terms in \eqref{sun1}). Though, it might not be clear at first sight, a simplification of the third term in \eqref{sun1} is still possible. We proceed as follows:
\begin{align}\label{sun2}
& \int \frac{d^d{p}}{(2\pi)^d}~\frac{d^{d}{q}}{(2\pi)^d} \sum_{l_{p}\neq 0}\frac{e^{i l_{p}.(\vec{p}+\vec{q})}}{(\vec{p}^2+m^2)(\vec{q}^2+m^2)((\vec{p}+\vec{q})^2+m^2)}\nonumber\\
& =  \int \frac{d^d{p}'}{(2\pi)^d}~\frac{d^{d}{q}'}{(2\pi)^d}\sum_{l_{p}\neq 0} \frac{e^{i l_{p}.\sqrt{2}\vec{p}'}}{\left(2\vec{p'}^2+m^2\right)\left(\frac{1}{2}\left(\vec{p'}-\vec{q'}\right)^2+m^2\right)\left(\frac{1}{2}\left(\vec{p'}+\vec{q'}\right)^2+m^2\right)}\nonumber\\
& = \int \frac{d^d{p}'}{(2\pi)^d}\sum_{l_{p}\neq 0}\frac{e^{i l_{p}.\sqrt{2}\vec{p}'}}{\left(2\vec{p'}^2+m^2\right)}\int \frac{d^d{q}'}{(2\pi)^d}\int_{0}^{1}dx\frac{1}{\left[\left(\frac{1}{2}\left(\vec{p'}-\vec{q'}\right)^2+m^2\right)x+\left(\frac{1}{2}\left(\vec{p'}+\vec{q'}\right)^2+m^2\right)(1-x)\right]^2}\nonumber\\
& =  \int \frac{d^d{p}'}{(2\pi)^d}\sum_{l_{p}\neq 0}\frac{e^{i l_{p}.\sqrt{2}\vec{p}'}}{\left(2\vec{p'}^2+m^2\right)}\int \frac{d^d{\tilde{q}}'}{(2\pi)^d}\int_{0}^{1}dx\frac{1}{\left[\frac{\vec{\tilde{q}}'^2}{2}+2\vec{p '}^2x(1-x)+m^2\right]^2}\nonumber\\
& =  \int \frac{d^d{\tilde{p}}}{(2\pi)^d}\sum_{l_{p}\neq 0}\frac{e^{i l_{p}.\vec{\tilde{p}}}}{\left(\vec{\tilde{p}}^2+m^2\right)}\int \frac{d^d{\tilde{q}}}{(2\pi)^d}\int_{0}^{1}dx\frac{1}{\left[\vec{\tilde{q}}^2+\vec{\tilde{p}}^2x(1-x)+m^2\right]^2}\nonumber\\
& = \int \frac{d^d{p}}{(2\pi)^d}~\frac{d^{d}{q}}{(2\pi)^d}\sum_{l_{p}\neq 0}\frac{e^{i l_{p}.\vec{p}}}{(\vec{p}^2+m^2)(\vec{q}^2+m^2)((\vec{p}+\vec{q})^2+m^2)}.
\end{align}
  Using \eqref{sun2} in \eqref{sun1}, we get :
  \begin{align}\label{sun3}
& \frac{1}{L^{2d}}\sum_{\vec{p},\vec{q}}\frac{1}{(\vec{p}^2+m^2)(\vec{q}^2+m^2)((\vec{p}+\vec{q})^2+m^2)}\nonumber\\
& = \int \frac{d^d{p}}{(2\pi)^d}~\frac{d^{d}{q}}{(2\pi)^d}\frac{1}{(\vec{p}^2+m^2)(\vec{q}^2+m^2)((\vec{p}+\vec{q})^2+m^2)}\nonumber\\
& +3 \int \frac{d^d{p}}{(2\pi)^d}~\frac{d^{d}{q}}{(2\pi)^d}\sum_{l_{p}\neq 0}\frac{e^{i l_{p}.\vec{p}}}{(\vec{p}^2+m^2)(\vec{q}^2+m^2)((\vec{p}+\vec{q})^2+m^2)}\nonumber\\
& + \int \frac{d^d{p}}{(2\pi)^d}~\frac{d^{d}{q}}{(2\pi)^d}\sum_{\substack{\{{l_{p}},{l_{q}}\}\neq 0,\\ {l_{p}}\neq {l_{q}}}}\frac{e^{i l_{p}.\vec{p}+i l_{q}.\vec{q}}}{(\vec{p}^2+m^2)(\vec{q}^2+m^2)((\vec{p}+\vec{q})^2+m^2)}.
\end{align}
An important comment is again in order. The first term in \eqref{sun3} i.e. the contribution from infinite volume contains both local and non-local divergences.The second term in \eqref{sun3} just contains non-local divergences, while the last term in \eqref{sun3} is free of any divergences. The first term of \eqref{sun3} is given by the following relation \cite{chung1997three, Two-loop, Two-loop-two-point}:
\begin{equation}\label{sun4p}
 \int \frac{d^d{p}}{(2\pi)^d}~\frac{d^{d}{q}}{(2\pi)^d}\frac{1}{(\vec{p}^2+m^2)(\vec{q}^2+m^2)((\vec{p}+\vec{q})^2+m^2)} = \frac{m^2}{2(4\pi)^4}\left(\frac{m^2}{4\pi}\right)^{-\epsilon}\frac{\Gamma^{2}\left(1+\frac{\epsilon}{2}\right)}{\left(1-\frac{\epsilon}{2}\right)\left(1-\epsilon\right)}\left[-\frac{12}{\epsilon^2}-6A+O(\epsilon)\right],
\end{equation}
with :
\begin{equation}
A = -\frac{2}{\sqrt{3}}\int_{0}^{\frac{\pi}{3}}d\theta~\log{\left(2 \sin{\frac{\theta}{2}}\right)}.
\end{equation}
Now, we set out to compute the second term in \eqref{sun3} which has both convergent as well as divergent parts in $d=4-\epsilon$. The calculation once again follows akin to \cite{bijnens2014two}:
\begin{align}\label{sun4}
 & \int \frac{d^d{p}}{(2\pi)^d}~\frac{d^{d}{q}}{(2\pi)^d}\sum_{l_{p}\neq 0}\frac{e^{i l_{p}.\vec{p}}}{(\vec{p}^2+m^2)(\vec{q}^2+m^2)((\vec{p}+\vec{q})^2+m^2)}\nonumber\\
 & = \int \frac{d^d{{p}}}{(2\pi)^d}\sum_{l_{p}\neq 0}\frac{e^{i l_{p}.\vec{p}}}{\left(\vec{{p}}^2+m^2\right)}\int \frac{d^d{q}}{(2\pi)^d}\int_{0}^{1}dx\frac{1}{\left[\vec{{q}}^2+\vec{{p}}^2x(1-x)+m^2\right]^2}\nonumber\\
 & = \int \frac{d^d{{p}}}{(2\pi)^d}\sum_{l_{p}\neq 0}\frac{e^{i l_{p}.\vec{p}}}{\left(\vec{{p}}^2+m^2\right)}\int_{0}^{1}\frac{dx}{(4\pi)^{d/2}}\Gamma\left[2-\frac{d}{2}\right]\left(x(1-x)\vec{p}^2+m^2\right)^{d/2-2}\nonumber\\
 & =  \int \frac{d^d{{p}}}{(2\pi)^d}\sum_{l_{p}\neq 0}\frac{e^{i l_{p}.\vec{p}}}{\left(\vec{{p}}^2+m^2\right)}\int_{0}^{1}\frac{dx}{(4\pi)^2}(4\pi)^{\epsilon/2}\left(\frac{2}{\epsilon}-\gamma_{E}+O(\epsilon)\right)\left(1-\frac{\epsilon}{2}\log{\{x(1-x)\vec{p}^2+m^2\}}+O(\epsilon)\right)\nonumber\\
 & =  \int \frac{d^d{{p}}}{(2\pi)^d}\sum_{l_{p}\neq 0}\frac{e^{i l_{p}.\vec{p}}}{\left(\vec{{p}}^2+m^2\right)}\int_{0}^{1}\frac{dx}{16\pi^2} \left[\frac{2}{\epsilon}+\log{4\pi}-\gamma_{E}-\log{\{x(1-x)\vec{p}^2+m^2\}}+O(\epsilon)\right]\nonumber\\
 & = \frac{2}{16\pi^2\epsilon} \int \frac{d^d{{p}}}{(2\pi)^d}\sum_{l_{p}\neq 0}\frac{e^{i l_{p}.\vec{p}}}{\left(\vec{{p}}^2+m^2\right)}+\frac{1}{16\pi^2}\left(\log{4\pi}-\gamma_{E}-\log{m^2}\right)\int \frac{d^d{{p}}}{(2\pi)^d}\sum_{l_{p}\neq 0}\frac{e^{i l_{p}.\vec{p}}}{\left(\vec{{p}}^2+m^2\right)}\nonumber\\
 & +\int \frac{d^d{{p}}}{(2\pi)^d}\sum_{l_{p}\neq 0}\frac{e^{i l_{p}.\vec{p}}}{\left(\vec{{p}}^2+m^2\right)}\int_{0}^{1}\frac{dx}{16\pi^2}~\frac{x(1-2x)\vec{p}^2}{x(1-x)\vec{p}^2+m^2}+O(\epsilon).
\end{align}
In \cite{sahu2024generalization}, it has already been shown that:
\begin{equation}\label{1Ldiscs}
    \frac{1}{L^d}\sum_{q\neq 0}\log{\left(1+\frac{r}{q^2}\right)} = \frac{1}{L^d}\left(\theta\left(L^2r /4\pi\right)-\theta(0)\right)+\int \frac{d^{d}k}{(2\pi)^d}\log\left(1+\frac{r}{k^2}\right),
\end{equation}
with:
\begin{equation}
\theta(z)=-\int_0^\infty d\sigma \frac{e^{-\sigma z}}\sigma\left(\vartheta^d(\sigma)-1-\sigma^{-d/2}\right).
\end{equation}
where $\vartheta(\sigma)$ is the Jacobi-theta function given by: $\vartheta(\sigma)=\sum_{j=-\infty}^{\infty}e^{-j^2 \pi \sigma}$.\\
Derivating it $n$ times with respect to $r$ one hence obtains:
\begin{equation}\label{discrete}
    \frac{1}{L^d}\sum_{q\neq 0}\frac{1}{\left(q^2+r\right)^n} = \frac{(-1)^{n-1}}{(n-1)!}\frac{1}{L^d}\left(\frac{L^2}{4\pi}\right)^n\theta^{(n)}\left(\frac{L^2r}{4\pi}\right)+\int \frac{d^{d}k}{(2\pi)^d}\frac{1}{\left(k^2+r\right)^n},
\end{equation}
where $\theta^{(n)}\left(m\right)$ means taking the derivative of $\theta\left(m\right)$, $n$ times with respect to $m$.
Putting $n=1$ in \eqref{discrete}, we have:
\begin{equation}\label{discrete1}
     \frac{1}{L^d}\sum_{q\neq 0}\frac{1}{\left(q^2+r\right)} = \frac{1}{L^d}\left(\frac{L^2}{4\pi}\right)\theta^{(1)}\left(\frac{L^2r}{4\pi}\right)+\int \frac{d^{d}k}{(2\pi)^d}\frac{1}{\left(k^2+r\right)}.
\end{equation}
Once again using Poisson summation rule, one can obtain the following relation:
\begin{equation}\label{sumP}
    \frac{1}{L^d}\sum_{q\neq 0}\frac{1}{\left(q^2+r\right)} +\frac{1}{L^d r} = \frac{1}{(2\pi)^d}\int \frac{d^{d}q}{\vec{q}^2+r}+\int \frac{d^{d}q}{(2\pi)^d}\sum_{l_{q}\neq 0}\frac{e^{il_{q}.\vec{q}}}{\vec{q}^2+r}. 
\end{equation}
Using\eqref{sumP} and \eqref{discrete1}, one hence obtains:
\begin{equation}\label{discrete2}
    \frac{1}{(2\pi)^d}\int d^{d}p\sum_{l_{p}\neq 0}\frac{e^{il_{p}.\vec{p}}}{\vec{p}^2+r} = \frac{1}{L^d}\left[\frac{L^2}{4\pi}\theta^{(1)}\left(\frac{L^2r}{4\pi}\right)+\frac{1}{r}\right]
\end{equation}

Using \eqref{discrete2} in \eqref{sun4}, we obtain:
\begin{align}\label{sunx}
    & \int \frac{d^d{p}}{(2\pi)^d}~\frac{d^{d}{q}}{(2\pi)^d}\sum_{l_{p}\neq 0}\frac{e^{i l_{p}.\vec{p}}}{(\vec{p}^2+m^2)(\vec{q}^2+m^2)((\vec{p}+\vec{q})^2+m^2)}\nonumber\\
    & = \frac{1}{16\pi^2}\left[\frac{2}{\epsilon}+\left(\log{4\pi}-\gamma_{E}-\log{m^2}\right)\right]\frac{1}{L^d}\left(\frac{L^2}{4\pi}\theta^{(1)}\left(\frac{m^2 L^2}{4\pi}\right)+\frac{1}{m^2}\right)+\int \frac{d^d{{p}}}{(2\pi)^d}\sum_{l_{p}\neq 0}\frac{e^{i l_{p}.\vec{p}}}{\left(\vec{{p}}^2+m^2\right)}\int_{0}^{1}\frac{dx}{16\pi^2}~\frac{x(1-2x)\vec{p}^2}{x(1-x)\vec{p}^2+m^2}.
\end{align}
In \eqref{sunx}, the term $\frac{2}{16\pi^2 \epsilon}\frac{1}{L^{d-2}}\left(\theta^{(1)}\left(\frac{m^2 L^2}{4\pi}\right)+\frac{1}{m^2 L^2}\right)$ is the non local divergence that in principle should cancel out in the renormalization procedure. We show how this happens in section \ref{2loopcal}. This also serves as a check for the perturbative renormalizability of the model under consideration, therefore, making sure that our final expression is actually free of any UV divergences in $d=4-\epsilon$. We now turn our attention to the computation of the last term in \eqref{sunx}.
\begin{align}\label{sunx1}
& \int \frac{d^d{{p}}}{(2\pi)^d}\sum_{l_{p}\neq 0}\frac{e^{i l_{p}.\vec{p}}}{\left(\vec{{p}}^2+m^2\right)}\int_{0}^{1}\frac{dx}{16\pi^2}~\frac{x(1-2x)\vec{p}^2}{x(1-x)\vec{p}^2+m^2}\nonumber\\
& = \frac{1}{16\pi^2}\int \frac{d^d{{p}}}{(2\pi)^d}\sum_{l_{p}\neq 0}\int_{0}^{\infty}d\alpha\int_{0}^{\infty}d\beta' ~e^{i l_{p}.\vec{p}-\alpha(\vec{p}^2+m^2)}\int_{0}^{1} dx~x(1-2x)\vec{p}^2 e^{-\beta'[x(1-x)\vec{p}^2+m^2]}\nonumber\\
& =\frac{1}{16\pi^2}\int_{0}^{\infty}d\alpha\int_{0}^{\infty}d\beta'\int_{0}^{1}dx\sum_{l_{p}\neq 0}\int\frac{d^{d}\tilde{p}}{(2\pi)^d}\frac{e^{-\vec{\tilde{p}}^2-\frac{l^{2}_{p}}{4[\alpha+\beta' x(1-x)]}}}{\left[\alpha+\beta' x(1-x)\right]^{\frac{d}{2}+1}}x(1-2x)\left[\vec{\tilde{p}}^2-\frac{l^{2}_{p}}{4(\alpha+\beta' x(1-x))}\right]e^{-m^2(\alpha+\beta')}\nonumber\\
& = \frac{1}{16\pi^2}\frac{\pi^{\frac{d}{2}}}{(2\pi)^d}\int_{0}^{\infty}d\alpha\int_{0}^{\infty}d\beta'\int_{0}^{1}dx\sum_{l_{p}\neq 0}e^{-\frac{l^{2}_{p}}{4[\alpha+\beta' x(1-x)]}}\frac{x(1-2x)e^{-m^{2}(\alpha+\beta')}}{\left[\alpha+\beta' x(1-x)\right]^{\frac{d}{2}+1}}\left[\frac{d}{2}-\frac{l^{2}_{p}}{4(\alpha+\beta' x(1-x))}\right]
\end{align}
In going from line 2 to line 3 in \eqref{sunx1}, we have made the following change of variable: $\vec{\tilde{p}}=\vec{p}\sqrt{\alpha+\beta x(1-x)}-\frac{il_{p}}{2\sqrt{\alpha+\beta x(1-x)}}$. In going from line 3 to line 4 in \eqref{sunx1}, we have integrate over all the $\tilde{p}$ modes. \eqref{sunx1} can once again be simplified using the following change of variables:$\gamma=\beta' x$, $\beta=\beta'(1-x)$. Thus \eqref{sunx1}, now becomes:
\begin{align}\label{sunx2}
   & \int \frac{d^d{{p}}}{(2\pi)^d}\sum_{l_{p}\neq 0}\frac{e^{i l_{p}.\vec{p}}}{\left(\vec{{p}}^2+m^2\right)}\int_{0}^{1}\frac{dx}{16\pi^2}~\frac{x(1-2x)\vec{p}^2}{x(1-x)\vec{p}^2+m^2}\nonumber\\
   & = \frac{1}{16\pi^2}\frac{1}{(4\pi)^{\frac{d}{2}}}\int_{0}^{\infty}d\alpha\int_{0}^{\infty}d\beta \int_{0}^{\infty}d\gamma\frac{(\beta+\gamma)^{\frac{d}{2}-2}e^{-(\alpha+\beta+\gamma)m^2}}{(\alpha\beta+\beta\gamma+\gamma\alpha)^{\frac{d}{2}+1}}\gamma(\beta-\gamma)\sum_{l_{p}\neq 0}e^{-\frac{l^{2}_{p}(\beta+\gamma)}{4[\alpha\beta+\beta\gamma+\gamma\alpha]}}\left(\frac{d}{2}-\frac{l^{2}_{p}(\beta+\gamma)}{4[\alpha\beta+\beta\gamma+\gamma\alpha]}\right)\nonumber\\
   & = \frac{1}{16\pi^2}\frac{1}{(4\pi)^{\frac{d}{2}}}\int_{0}^{\infty}d\alpha\int_{0}^{\infty}d\beta \int_{0}^{\infty}d\gamma\frac{(\beta+\gamma)^{\frac{d}{2}-2}e^{-(\alpha+\beta+\gamma)m^2}}{(\alpha\beta+\beta\gamma+\gamma\alpha)^{\frac{d}{2}+1}}\gamma(\beta-\gamma)\nonumber\\
   & \times\left[\frac{d}{2}\left(\vartheta^{d}\left(\frac{(\beta+\gamma)L^2}{4\pi\left[\alpha\beta+\beta\gamma+\gamma\alpha\right]}\right)-1\right)+\frac{(\beta+\gamma)L^2 }{4\pi\left[\alpha\beta+\beta\gamma+\gamma\alpha\right]}d~\vartheta^{d-1}\left(\frac{(\beta+\gamma)L^2}{4\pi\left[\alpha\beta+\beta\gamma+\gamma\alpha\right]}\right)\vartheta'\left(\frac{(\beta+\gamma)L^2}{4\pi\left[\alpha\beta+\beta\gamma+\gamma\alpha\right]}\right)\right].
\end{align}
where $\vartheta'(\sigma)$ represents the derivative of $\vartheta$ with respect to $\sigma$. It is also easy to check that \eqref{sunx2} is actually convergent in $d=4-\epsilon$ and the only divergence thus appears in the first term of \eqref{sunx}. One must also not forget that $\theta^{(1)}$ itself is also a dimension dependent quantity and hence in principle one should also expand $\theta^{(1)}$ in an epsilon expansion. However it doesn't really matter since the entirety of $\frac{2}{16\pi^2\epsilon}\theta^{(1)}\left(\frac{m^2 L^2}{4\pi}\right)$ actually cancels due to the counterterms introduced as a result of the renormalization prescription.
Hence \eqref{sunx} thus becomes:
\begin{align}\label{sunxf}
   & \int \frac{d^d{p}}{(2\pi)^d}~\frac{d^{d}{q}}{(2\pi)^d}\sum_{l_{p}\neq 0}\frac{e^{i l_{p}.\vec{p}}}{(\vec{p}^2+m^2)(\vec{q}^2+m^2)((\vec{p}+\vec{q})^2+m^2)}\nonumber\\
    & = \frac{1}{16\pi^2}\left[\frac{2}{\epsilon}+\left(\log{4\pi}-\gamma-\log{m^2}\right)\right]\frac{1}{L^d}\left(\frac{L^2}{4\pi}\theta^{(1)}\left(\frac{m^2 L^2}{4\pi}\right)+\frac{1}{m^2}\right)\nonumber\\
    &+  \frac{1}{16\pi^2}\frac{1}{(4\pi)^{\frac{d}{2}}}\int_{0}^{\infty}d\alpha\int_{0}^{\infty}d\beta \int_{0}^{\infty}d\gamma\frac{(\beta+\gamma)^{\frac{d}{2}-2}e^{-(\alpha+\beta+\gamma)m^2}}{(\alpha\beta+\beta\gamma+\gamma\alpha)^{\frac{d}{2}+1}}\gamma(\beta-\gamma)\nonumber\\
   & \times\left[\frac{d}{2}\left(\vartheta^{d}\left(\frac{(\beta+\gamma)L^2}{4\pi\left[\alpha\beta+\beta\gamma+\gamma\alpha\right]}\right)-1\right)+\frac{(\beta+\gamma)L^2 }{4\pi\left[\alpha\beta+\beta\gamma+\gamma\alpha\right]}d~\vartheta^{d-1}\left(\frac{(\beta+\gamma)L^2}{4\pi\left[\alpha\beta+\beta\gamma+\gamma\alpha\right]}\right)\vartheta'\left(\frac{(\beta+\gamma)L^2}{4\pi\left[\alpha\beta+\beta\gamma+\gamma\alpha\right]}\right)\right].
\end{align}
Now we turn our attention to computing the last term in \eqref{sun}:
\begin{align}\label{sunxN}
   & \int \frac{d^d{p}}{(2\pi)^d}~\frac{d^{d}{q}}{(2\pi)^d}\sum_{\substack{\{{l_{p}},{l_{q}}\}\neq 0,\\ {l_{p}}\neq {l_{q}}}}\frac{e^{i l_{p}.\vec{p}+i l_{q}.\vec{q}}}{(\vec{p}^2+m^2)(\vec{q}^2+m^2)((\vec{p}+\vec{q})^2+m^2)}\nonumber\\
   & = \int \frac{d^d{p}}{(2\pi)^d}~\frac{d^{d}{q}}{(2\pi)^d}\sum_{\substack{\{{l_{p}},{l_{q}}\}\neq 0,\\ {l_{p}}\neq {l_{q}}}}\int_{0}^{\infty}d\alpha\int_{0}^{\infty}d\beta\int_{0}^{\infty}d\gamma ~e^{-(\alpha+\gamma)\vec{p}^2-(\beta+\gamma)\vec{q}^2-2\gamma \vec{p}.\vec{q}+il_{p}.\vec{p}+il_{q}.\vec{q}-(\alpha+\beta+\gamma)m^2}\nonumber\\
   & = \int \frac{d^d{p}}{(2\pi)^d}~\frac{d^{d}{q}}{(2\pi)^d}\sum_{\substack{\{{l_{p}},{l_{q}}\}\neq 0,\\ {l_{p}}\neq {l_{q}}}}\int_{0}^{\infty}d\alpha\int_{0}^{\infty}d\beta\int_{0}^{\infty}d\gamma
   ~e^{-\begin{pmatrix}
       \vec{p} & \vec{q}
   \end{pmatrix}\begin{pmatrix}
   \alpha+\gamma & \gamma\\
   \gamma & \beta+\gamma
   \end{pmatrix}\begin{pmatrix}
       \vec{p}\\
       \vec{q}
   \end{pmatrix}}e^{il_{p}.\vec{p}+il_{q}.\vec{q}-(\alpha+\beta+\gamma)m^2}.
\end{align}
We choose to work in a rotated frame parametrized by co-ordinates $\vec{p'}$, $\vec{q'}$, given by the following transformation:
$$\begin{pmatrix}
    \vec{p}\\
    \vec{q}
\end{pmatrix}=\begin{pmatrix}
    \sqrt{\frac{\beta+\gamma}{\alpha\beta+\beta\gamma+\gamma\alpha}} & 0\\
    -\frac{\gamma}{\sqrt{(\beta+\gamma)(\alpha\beta+\beta\gamma+\gamma\alpha)}} & \frac{1}{\sqrt{\beta+\gamma}}
\end{pmatrix}\begin{pmatrix}
        \vec{p'}\\
        \vec{q'}.
    \end{pmatrix}$$
    Thus \eqref{sunxN} becomes:
    \begin{align}\label{sunxN1}
          & \int \frac{d^d{p}}{(2\pi)^d}~\frac{d^{d}{q}}{(2\pi)^d}\sum_{\substack{\{{l_{p}},{l_{q}}\}\neq 0,\\ {l_{p}}\neq {l_{q}}}}\frac{e^{i l_{p}.\vec{p}+i l_{q}.\vec{q}}}{(\vec{p}^2+m^2)(\vec{q}^2+m^2)((\vec{p}+\vec{q})^2+m^2)}\nonumber\\
          & = \int \frac{d^d{p'}}{(2\pi)^d}~\frac{d^{d}{q'}}{(2\pi)^d}\sum_{\substack{\{{l_{p}},{l_{q}}\}\neq 0,\\ {l_{p}}\neq {l_{q}}}}\int_{0}^{\infty}d\alpha\int_{0}^{\infty}d\beta\int_{0}^{\infty}d\gamma~\frac{1}{(\alpha\beta+\beta\gamma+\gamma\alpha)^{d/2}} e^{-\vec{p'}^2-\vec{q'}^2+i l_{p}.\vec{p'}\sqrt{\frac{\beta+\gamma}{\alpha\beta+\beta\gamma+\gamma\alpha}}+il_{q}.\left(\frac{\vec{q'}}{\sqrt{\beta+\gamma}}-\frac{\gamma\vec{p'}}{\sqrt{(\beta+\gamma)(\alpha\beta+\beta\gamma+\gamma\alpha)}}\right)-(\alpha+\beta+\gamma)m^2}.
    \end{align}
Making the following change of variables: 
\begin{equation}
    \vec{\tilde{p}} = \vec{p'}-\frac{i l_{p}}{2}\sqrt{\frac{\beta+\gamma}{\alpha\beta+\beta\gamma+\gamma\alpha}}+\frac{i\gamma l_{q}}{2\sqrt{(\beta+\gamma)(\alpha\beta+\beta\gamma+\gamma\alpha)}}\ , \ \ \vec{\tilde{q}}=\vec{q'}-\frac{il_{q}}{2\sqrt{\beta+\gamma}},
\end{equation}
\eqref{sunxN1} now becomes:
\begin{align}\label{sunxN2}
    & \int \frac{d^d{p}}{(2\pi)^d}~\frac{d^{d}{q}}{(2\pi)^d}\sum_{\substack{\{{l_{p}},{l_{q}}\}\neq 0,\\ {l_{p}}\neq {l_{q}}}}\frac{e^{i l_{p}.\vec{p}+i l_{q}.\vec{q}}}{(\vec{p}^2+m^2)(\vec{q}^2+m^2)((\vec{p}+\vec{q})^2+m^2)}\nonumber\\
    & = \int \frac{d^d{\tilde {p}}}{(2\pi)^d}~\frac{d^{d}{\tilde{q}}}{(2\pi)^d}\sum_{\substack{\{{l_{p}},{l_{q}}\}\neq 0,\\ {l_{p}}\neq {l_{q}}}}\int_{0}^{\infty}d\alpha\int_{0}^{\infty}d\beta\int_{0}^{\infty}d\gamma\frac{1}{(\alpha\beta+\beta\gamma+\gamma\alpha)^{d/2}}e^{-\vec{\tilde{p}}^2 - \vec{\tilde{q}}^2-(\alpha+\beta+\gamma)m^2-\frac{1}{4(\alpha\beta+\beta\gamma+\gamma\alpha)}[\beta l^{2}_{p}+\alpha l^2_{q}+\gamma(l_{p}-l_{q})^2]}\nonumber\\
    & = \frac{1}{(4\pi)^d}\int_{0}^{\infty}d\alpha\int_{0}^{\infty}d\beta\int_{0}^{\infty}d\gamma\frac{e^{-(\alpha+\beta+\gamma)m^2}}{(\alpha\beta+\beta\gamma+\gamma\alpha)^{d/2}}\sum_{\substack{\{{l_{p}},{l_{q}}\}\neq 0,\\ {l_{p}}\neq {l_{q}}}}e^{-\frac{1}{4(\alpha\beta+\beta\gamma+\gamma\alpha)}[\beta l^{2}_{p}+\alpha l^2_{q}+\gamma(l_{p}-l_{q})^2]}.
\end{align}
We introduce the Riemann-theta function $\vartheta^{(g)}(z|\tau)= \sum_{n\in Z^{g}}e^{2\pi i (\frac{1}{2} n^{T}.\tau.n+n^{T}.z)}$ whose generalization in 1-dimension happens to be the Jacobi theta function defined earlier. In terms of these theta functions \eqref{sunxN2}, can be written as:
\begin{align}\label{sunxN3}
    \int \frac{d^d{p}}{(2\pi)^d}~\frac{d^{d}{q}}{(2\pi)^d}\sum_{\substack{\{{l_{p}},{l_{q}}\}\neq 0,\\ {l_{p}}\neq {l_{q}}}}&\frac{e^{i l_{p}.\vec{p}+i l_{q}.\vec{q}}}{(\vec{p}^2+m^2)(\vec{q}^2+m^2)((\vec{p}+\vec{q})^2+m^2)}\nonumber\\
     = \frac{1}{(4\pi)^d}\int_{0}^{\infty}d\alpha\int_{0}^{\infty}d\beta\int_{0}^{\infty}d\gamma\frac{e^{-(\alpha+\beta+\gamma)m^2}}{\mathcal{S}^{d/2}} &\biggl[\left(\vartheta^{(2)}\left(0|\frac{i L^2}{4\pi \mathcal{S}}\begin{bmatrix}
   \beta+\gamma & -\gamma\\
   -\gamma & \alpha+\gamma
   \end{bmatrix}\right)\right)^{d}-\vartheta^{d}\left(\frac{(\alpha+\beta)L^2}{4\pi\mathcal{S}}\right)\nonumber\\
    &-\vartheta^{d}\left(\frac{(\beta+\gamma)L^2}{4\pi\mathcal{S}}\right)-\vartheta^{d}\left(\frac{(\gamma+\alpha)L^2}{4\pi\mathcal{S}}\right)+2\biggr].
\end{align}

Now, rescaling the dummy variables $\alpha$, $\beta$, and $\gamma$ by $L^{2}$ in both \eqref{sunxN3} and \eqref{sunxf}, and using \eqref{sun4p} one obtains in $d=4-\epsilon$:
\begin{align}
    & \frac{1}{L^{2d}}\sum_{\vec{p},\vec{q}}\frac{1}{(\vec{p}^2+m^2)(\vec{q}^2+m^2)((\vec{p}+\vec{q})^2+m^2)}\nonumber\\
    & =\frac{1}{L^{2d-6}}\biggl[ \frac{1}{(16\pi^2)^2}\int_{0}^{\infty}d\alpha\int_{0}^{\infty}d\beta\int_{0}^{\infty}d\gamma\frac{e^{-(\alpha+\beta+\gamma)m^2 L^2}}{\mathcal{S}^{2}} \biggl[\left(\vartheta^{(2)}\left(0|\frac{i}{4\pi \mathcal{S}}\begin{bmatrix}
   \beta+\gamma & -\gamma\\
   -\gamma & \alpha+\gamma
   \end{bmatrix}\right)\right)^{4}-\vartheta^{4}\left(\frac{(\alpha+\beta)}{4\pi\mathcal{S}}\right)\nonumber\\
    &-\vartheta^{4}\left(\frac{(\beta+\gamma)}{4\pi\mathcal{S}}\right)-\vartheta^{4}\left(\frac{(\gamma+\alpha)}{4\pi\mathcal{S}}\right)+2\biggr] + \frac{3}{(16\pi^2)^2}\int_{0}^{\infty}d\alpha\int_{0}^{\infty}d\beta \int_{0}^{\infty}d\gamma\frac{e^{-(\alpha+\beta+\gamma)m^2 L^2}}{\mathcal{S}^{3}}\gamma(\beta-\gamma)\nonumber\\
   & \times\left[2\left(\vartheta^{4}\left(\frac{(\beta+\gamma)}{4\pi\mathcal{S}}\right)-1\right)+4\frac{(\beta+\gamma) }{4\pi\mathcal{S}}~\vartheta^{3}\left(\frac{(\beta+\gamma)}{4\pi\mathcal{S}}\right)\vartheta'\left(\frac{(\beta+\gamma)}{4\pi\mathcal{S}}\right)\right]+\frac{3}{16\pi^2}\left(\log{4\pi}-\gamma_{E}-\log{m^2 L^2}\right)\frac{1}{4\pi}\theta^{(1)}\left(\frac{m^2 L^2}{4\pi}\right)\nonumber\\
   &+\frac{3}{16\pi^2}\left(\log{4\pi}-\gamma_{E}-\log{m^2 L^2}\right)\frac{1}{m^2 L^2} +\frac{3}{16\pi^2}\frac{2}{\epsilon}\left(\frac{1}{4\pi}\theta^{(1)}\left(\frac{m^2 L^2}{4\pi}\right)+\frac{1}{m^2 L^2}\right)+ \frac{m^2 L^2}{(16\pi^2)^2}\biggl[-\frac{6}{\epsilon^2}\nonumber\\
   & +\frac{6}{\epsilon}\left(\gamma_{E}-\frac{3}{2}+\log{\left(\frac{m^2 L^2}{4\pi}\right)}\right)-3 A-\frac{21}{2}+9\gamma_{E}-3\gamma_{E}^2-\frac{\pi^2}{4}-(6\gamma_{E}-9)\log{\left(\frac{m^2 L^2}{4\pi}\right)}-3\log^{2}{\left(\frac{m^2 L^2}{4\pi}\right)}\biggr]+O(\epsilon)\biggr].
\end{align}
with $\gamma_{E}$, the Euler-Mascheroni constant.\\
Using \eqref{discrete} in \eqref{sun}, one thereby obtains in $d=4-\epsilon$:
\begin{align}\label{sunf}
 & \frac{1}{L^{2d}}\sum_{\substack{\{\vec{p},\vec{q}\}\neq 0,\\ \vec{p}\neq -\vec{q}}}\frac{1}{(\vec{p}^2+m^2)(\vec{q}^2+m^2)((\vec{p}+\vec{q})^2+m^2)}\nonumber\\
    & =\frac{1}{L^{2d-6}}\biggl[ \frac{1}{(16\pi^2)^2}\int_{0}^{\infty}d\alpha\int_{0}^{\infty}d\beta\int_{0}^{\infty}d\gamma\frac{e^{-(\alpha+\beta+\gamma)m^2 L^2}}{\mathcal{S}^{2}} \biggl[\left(\vartheta^{(2)}\left(0|\frac{i}{4\pi \mathcal{S}}\begin{bmatrix}
   \beta+\gamma & -\gamma\\
   -\gamma & \alpha+\gamma
   \end{bmatrix}\right)\right)^{4}-\vartheta^{4}\left(\frac{(\alpha+\beta)}{4\pi\mathcal{S}}\right)\nonumber\\
    &-\vartheta^{4}\left(\frac{(\beta+\gamma)}{4\pi\mathcal{S}}\right)-\vartheta^{4}\left(\frac{(\gamma+\alpha)}{4\pi\mathcal{S}}\right)+2\biggr] + \frac{3}{(16\pi^2)^2}\int_{0}^{\infty}d\alpha\int_{0}^{\infty}d\beta \int_{0}^{\infty}d\gamma\frac{e^{-(\alpha+\beta+\gamma)m^2 L^2}}{\mathcal{S}^{3}}\gamma(\beta-\gamma)\nonumber\\
   & \times\left[2\left(\vartheta^{4}\left(\frac{(\beta+\gamma)}{4\pi\mathcal{S}}\right)-1\right)+4\frac{(\beta+\gamma) }{4\pi\mathcal{S}}~\vartheta^{3}\left(\frac{(\beta+\gamma)}{4\pi\mathcal{S}}\right)\vartheta'\left(\frac{(\beta+\gamma)}{4\pi\mathcal{S}}\right)\right]+\frac{3}{16\pi^2}\left(\log{4\pi}-\gamma_{E}-\log{m^2 L^2}\right)\frac{1}{4\pi}\theta^{(1)}\left(\frac{m^2 L^2}{4\pi}\right)\nonumber\\
   &+\frac{3}{16\pi^2}\frac{1}{m^2 L^2} \theta^{(2)}\left(\frac{m^2 L^2}{4\pi}\right) +\frac{3}{16\pi^2}\frac{2}{\epsilon}\left(\frac{1}{4\pi}\theta^{(1)}\left(\frac{m^2 L^2}{4\pi}\right)\right)+ \frac{m^2 L^2}{(16\pi^2)^2}\biggl[-\frac{6}{\epsilon^2} +\frac{6}{\epsilon}\left(\gamma_{E}-\frac{3}{2}+\log{\left(\frac{m^2 L^2}{4\pi}\right)}\right)\nonumber\\
   &-3 A-\frac{21}{2}+9\gamma_{E}-3\gamma_{E}^2-\frac{\pi^2}{4}-(6\gamma_{E}-9)\log{\left(\frac{m^2 L^2}{4\pi}\right)}-3\log^{2}{\left(\frac{m^2 L^2}{4\pi}\right)}\biggr]-\frac{1}{(m^2 L^2)^3}+O(\epsilon)\biggr],   
\end{align}

with $\mathcal{S}=\alpha\beta+\beta\gamma+\gamma\alpha$,  as defined before.\\

Now we will show how to compute the double bubble like diagram, the computation of which is rather simple. Using \eqref{discrete1}, we have:
\begin{align}
    &\left(\frac{1}{L^d}\sum_{\vec{q}\neq 0}\frac{1}{\vec{q}^2+m^2}\right)^2\nonumber\\
    & = \left(\frac{1}{L^d}\left(\frac{L^2}{4\pi}\right)\theta^{(1)}\left(\frac{m^2 L^2}{4\pi}\right)+\int \frac{d^{d}k}{(2\pi)^d}\frac{1}{\left(k^2+m^2\right)}\right)^2.
\end{align}
Using:
\begin{equation}
    \int \frac{d^{d}k}{(2\pi)^d}\frac{1}{\left(k^2+m^2\right)} = \frac{1}{(4\pi)^{d/2}}\Gamma\left(1-\frac{d}{2}\right)(m^{2})^{d/2-1},
\end{equation}
in $d=4-\epsilon$, we have:
\begin{align}\label{dbubblef}
    &\left(\frac{1}{L^d}\sum_{\vec{q}\neq 0}\frac{1}{\vec{q}^2+m^2}\right)^2\nonumber\\
    & = \frac{1}{L^{2d-4}}\biggl[\frac{1}{16\pi^2}\theta^{(1)}\left[\left(\frac{m^2 L^2}{4\pi}\right)\right]^2+\frac{2}{4\pi}\theta^{(1)}\left(\frac{m^2 L^2}{4\pi}\right)\frac{m^2 L^2}{(4\pi)^2}\left(\gamma_{E} -1+\log{\left(\frac{m^2 L^2}{4\pi}\right)}\right)+\frac{(m^2 L^2)^2}{(16\pi^2)^2}\biggl(2\gamma_{E}^2-4\gamma_{E}+3+\frac{\pi^2}{6}\nonumber\\
    & +4(\gamma_{E}-1)\log{\left(\frac{m^2 L^2}{4\pi}\right)}+2\log^{2}{\frac{m^2 L^2}{4\pi}}\biggr)+\frac{4}{\epsilon^2}\left(\frac{m^2 L^{2}}{16\pi^2}\right)^2-\frac{4}{16\pi^2\epsilon}\frac{m^2 L^2}{4\pi}\theta^{(1)}\left(\frac{m^2 L^2}{4\pi}\right)-\frac{4}{16\pi^2\epsilon}\frac{(m^2 L^2)^2}{(4\pi)^2}\biggl(\gamma_{E} -1\nonumber\\
    & +\log{\left(\frac{m^2 L^2}{4\pi}\right)}\biggr)+O(\epsilon)\biggr].
    \end{align}  

\section{}\label{appen D} 

This section contains some details of the calculation performed in Eq.~\eqref{divc-rate} and Eq. ~\eqref{conv-rate}. We start with 

\begin{equation}
    \lim_{M\rightarrow\infty}\Gamma_{M}=\Gamma_{MF}+\Gamma_{1L}+\Gamma_{2L},
\end{equation}
where, the $\tilde{s}$ dependence of $\Gamma_{M}$ has been suppressed in the above equation. Here $\Gamma_{MF}$ is the mean-field term,  $\Gamma_{1L}$ is the one-loop term and $\Gamma_{2L}$ is the two-loop term. One can easily identify $\Gamma_{MF}$, $\Gamma_{1L}$ and $\Gamma_{2L}$ from Eq.~\eqref{e1}. They are given by:

\begin{align}\label{GammaMF1}
    \Gamma_{MF} = \frac{1}{2}Z_{2}\tilde{t }\tilde{s}^2+\frac{1}{4!}Z_{4}\tilde{g} \tilde{s}^4,
\end{align}

\begin{align}
    \Gamma_{1L} = \frac{\hbar}{2}\sum_{\vec{q}\neq 0}\log\left(1+\frac{Z_{2}\tilde{t }+Z_{4}\tilde{s}^2/2}{\vec{q}^2}\right),
\end{align}
 and 
\begin{align}
    \Gamma_{2L}& =\frac{\tilde{g}}{8}\hbar^2\left(\sum_{\vec{q}\neq 0}\frac{1}{\vec{q}^2+\tilde{m}^2}\right)^2-\frac{\tilde{g}^2 \tilde{s}^2}{12}\hbar^2\sum_{\substack{\{\vec{p},\vec{q}\}\neq 0,\\ \vec{p}\neq -\vec{q}}}\frac{1}{(\vec{p}^2+\tilde{m}^2)(\vec{q}^2+\tilde{m}^2)((\vec{p}+\vec{q})^2+\tilde{m}^2)},
\end{align}
with : $\tilde{m}^2 = Z_{2}\tilde{t }+Z_{4}\tilde{s}^2/2$.\\

Obviously, $\Gamma_{MF}$, $\Gamma_{1L}$ and $\Gamma_{2L}$  defined above can be further broken down into their convergent and divergent halves. We represent the convergent and divergent halves of the quantities $\Gamma_{MF}$, $\Gamma_{1L}$ and $\Gamma_{2L}$ with the superscripts $(conv)$ and $(div)$ respectively. \\

Thus using Eq.~\eqref{counterterms} and Eq.\eqref{GammaMF1}, the mean-field term $\Gamma_{MF}$ is given by $\Gamma_{MF}=\Gamma^{(conv)}_{MF}+\Gamma^{(div)}_{MF}$ with :
\begin{align}
    \Gamma^{(conv)}_{MF} = \frac{1}{2}\tilde{t}\tilde{s}^2+\frac{1}{24}\tilde{g}\tilde{s}^4,
\end{align}

\begin{align}
      \Gamma^{(div)}_{MF} & = \frac{1}{2}\left(\frac{\tilde{g} \hbar}{16\pi^2\epsilon}+\frac{2\tilde{g}^{2}\hbar^2}{(16\pi^2)^2\epsilon^2}-\frac{\tilde{g}^2\hbar^2}{2(16\pi^2)^2\epsilon}\right)\tilde{t}\tilde{s}^{2}+\frac{1}{24}\left(\frac{3\tilde{g} \hbar}{16\pi^2\epsilon}+\frac{9\tilde{g}^{2}\hbar^2}{(16\pi^2)^2\epsilon^2}-\frac{3\tilde{ g}^2\hbar^2}{(16\pi^2)^2\epsilon}\right)\tilde{g}\tilde{s}^{4}.
\end{align}
Following Eq.\eqref{1Ldiscs}, the one-loop term $\Gamma_{1L}$ is thus similarly given by  $\Gamma_{1L}=\Gamma^{(conv)}_{1L}+\Gamma^{(div)}_{1L}$ with :

\begin{align}\label{epsilon1L}
   \Gamma^{(conv)}_{1L} & = \frac{\hbar}{4(16\pi^2)}\left(\tilde{t}+\frac{\tilde{g}\tilde{s}^2}{2}\right)^{2} \left[\gamma_{E}-\frac{3}{2}+\log{\left(\frac{\tilde{t}+\tilde{g}\tilde{s}^2/2}{4\pi}\right)}\right] +\frac{\hbar}{2}\Delta\left(\frac{1}{4\pi}\left[\tilde{t}+\frac{\tilde{g}\tilde{s}^2}{2}\right]\right)\nonumber\\
    &-\frac{\epsilon\hbar}{2}\Delta^{(\epsilon)}\left(\frac{1}{4\pi}\left[\tilde{t}+\frac{\tilde{g}\tilde{s}^2}{2}\right]\right) -\frac{\epsilon\hbar}{2}\frac{1}{48}\left(\frac{\tilde{t}+\tilde{g}\tilde{s}^2/2}{4\pi}\right)^2\left[21- 18\gamma_{E} + 6 \gamma_{E}^2 + \pi^2 + 6 (2\gamma_{E}-3 )\log{\left(\frac{\tilde{t}+\tilde{g}\tilde{s}^2/2}{4\pi}\right)}+6 \log^{2}{\left(\frac{\tilde{t}+\tilde{g}\tilde{s}^2/2}{4\pi}\right)} \right]\nonumber\\ 
    & -\frac{\hbar^2}{8(16\pi^2)^2}\tilde{g}\left(\tilde{t}+\frac{3\tilde{g}\tilde{s}^2}{2}\right)\left(\tilde{t}+\frac{\tilde{g}\tilde{s}^2}{2}\right)\left(\frac{\pi^2}{6}+2-2\gamma_{E}+\gamma_{E}^2+2(\gamma_{E}-1)\log{\left(\frac{ \tilde{t}+\tilde{g}\tilde{s}^2/2}{4\pi}\right)}+\log^{2}{\left(\frac{ \tilde{t}+\tilde{g}\tilde{s}^2/2}{4\pi}\right)}\right)
\end{align}

and

\begin{align}
    \Gamma^{(div)}_{1L} & = -\frac{\hbar}{2(16\pi^2)\epsilon}\left(\tilde{t}+\frac{\tilde{g}\tilde{s}^2}{2}\right)^{2}+\frac{\hbar^2}{2(16\pi^2)^2\epsilon}\tilde{g}\left(\tilde{t}+\frac{3\tilde{g}\tilde{s}^2}{2}\right)\left(\tilde{t}+\frac{\tilde{g}\tilde{s}^2}{2}\right)\biggl[-\frac{2}{\epsilon}+\gamma_{E}-1+\log{\left(\frac{\tilde{t}+\tilde{g}\tilde{s}^2/2}{4\pi}\right)}\biggr]\nonumber\\
       & +\hbar^{2}\frac{\tilde{g}}{32\pi^{2}\epsilon}\left(\tilde{t}+\frac{3\tilde{g}\tilde{s}^{2}}{2}\right)\left\{\frac{1}{4\pi}\theta^{(1)}\left(\frac{\left(\tilde{t}+\tilde{g}\tilde{s}^2/2\right)}{4\pi}\right)\right\},       
\end{align}

where:
    \begin{align}
        \Delta(\tilde{m}^{2})&=\theta(\tilde{m}^{2})-\theta(0)\nonumber\\
        & = \int_0^\infty d\sigma \frac{1-e^{-\sigma \tilde{m}^2}}\sigma\left(\vartheta^4(\sigma)-1-\sigma^{-2}\right),
    \end{align}
    and
    \begin{equation}
        \Delta^{(\epsilon)}(\tilde{m}^{2})=\int_{0}^{\infty}d\sigma \frac{1-e^{-\tilde{m}^{2}\sigma}}{\sigma}\left(\vartheta^{4}(\sigma)\log{\vartheta(\sigma)}+\frac{\log{\sigma}}{2\sigma^{2}}\right).
    \end{equation}

    (The function $\Delta^{(\epsilon)}(\tilde{m}^2)$ is nothing but the function appearing as a result of $4-\epsilon$ expansion of the function  $\Delta(\tilde{m}^2)$ ).

 Here, the first line in Eq.~\eqref{epsilon1L} contains purely one-loop terms and contribute at the first order of the $\epsilon-$expansion. The next line in Eq.~\eqref{epsilon1L} contains terms due to the $\epsilon-$ expansion of the one-loop term and contribute to the second order of the $\epsilon-$expansion and are thereby required to guarantee the consistency of the $\epsilon-$expansion after a suitable rescaling of $\tilde{s}$ (with a $\sqrt{u_{*}}$, see section \ref{2loopcal} ) is performed. The last line in Eq.~\eqref{epsilon1L} contains terms that also contribute to the second order of the $\epsilon-$expansion and are due to the introduction of counter-term terms at one-loop that have a finite contribution at two-loops.\\

Using Eqs.\eqref{sunf} and \eqref{dbubblef} , one can now extensively compute the two-loop term $\Gamma_{2L}=\Gamma^{(conv)}_{2L}+\Gamma^{(div)}_{2L}$ with:

\begin{align}
    \Gamma^{(conv)}_{2L} &=  -\hbar^{2}\frac{\tilde{g}^{2}\tilde{s}^2}{12} \frac{\left(\tilde{t}+\tilde{g}\tilde{s}^2/2\right)}{(16\pi^2)^2}\left[-3 A-\frac{21}{2}+9\gamma_{E}-3\gamma_{E}^2-\frac{\pi^2}{4}-(6\gamma_{E}-9)\log{\left(\frac{\left(\tilde{t}+\tilde{g}\tilde{s}^2/2\right)}{4\pi}\right)}-3\log^{2}{\left(\frac{\left(\tilde{t}+\tilde{g}\tilde{s}^2/2\right)}{4\pi}\right)}\right]\nonumber\\
  & +\hbar^2\frac{\tilde{g}}{8}\left[\frac{\left(\tilde{t}+\tilde{g}\tilde{s}^2/2\right)^2}{(16\pi^2)^2}\left(2\gamma_{E}^2-4\gamma_{E}+3+\frac{\pi^2}{6}+4(\gamma_{E}-1)\log{\left(\frac{\left(\tilde{t}+\tilde{g}\tilde{s}^2/2\right)}{4\pi}\right)}+2\log^{2}{\left(\frac{\tilde{t}+\tilde{g}\tilde{s}^2/2}{4\pi}\right)}\right)\right]\nonumber\\
 & -\hbar^{2}\frac{\tilde{g}^{2}\tilde{s}^2}{12}\biggl[ I_{1}[\tilde{t}+\tilde{g}\tilde{s}^2/2]+I_{2}[\tilde{t}+\tilde{g}\tilde{s}^2/2]+\frac{3}{16\pi^2}\left(\log{4\pi}-\gamma_{E}-\log{\left(\tilde{t}+\tilde{g}\tilde{s}^2/2\right)}\right)\frac{1}{4\pi}\theta^{(1)}\left(\frac{\left(\tilde{t}+\tilde{g}\tilde{s}^2/2\right)}{4\pi}\right)\nonumber\\
   &+\frac{3}{16\pi^2}\frac{1}{\left(\tilde{t}+\tilde{g}\tilde{s}^2/2\right)} \theta^{(2)}\left(\frac{\left(\tilde{t}+\tilde{g}\tilde{s}^2/2\right)}{4\pi}\right)-\frac{1}{\left(\tilde{t}+\tilde{g}\tilde{s}^2/2\right)^3}\biggr]+\hbar^2\frac{\tilde{g}}{8}\biggl[\frac{1}{16\pi^2}\left[\theta^{(1)}\left(\frac{\left(\tilde{t}+\tilde{g}\tilde{s}^2/2\right)}{4\pi}\right)\right]^2\nonumber\\
    & +\frac{2}{4\pi}\theta^{(1)}\left(\frac{\left(\tilde{t}+\tilde{g}\tilde{s}^2/2\right)}{4\pi}\right)\frac{\left(\tilde{t}+\tilde{g}\tilde{s}^2/2\right)}{(4\pi)^2}\left\{\gamma_{E} -1+\log{\left(\frac{\left(\tilde{t}+\tilde{g}\tilde{s}^2/2\right)}{4\pi}\right)}\right\}\biggr],
\end{align}

and 

\begin{align}
    \Gamma^{(div)}_{2L} &= -\hbar^{2}\frac{\tilde{g}^{2}\tilde{s}^2}{12} \frac{\left(\tilde{t}+\tilde{g}\tilde{s}^2/2\right)}{(16\pi^2)^2}\left[-\frac{6}{\epsilon^2} +\frac{6}{\epsilon}\left\{\gamma_{E}-\frac{3}{2}+\log{\left(\frac{\left(\tilde{t}+\tilde{g}\tilde{s}^2/2\right)}{4\pi}\right)}\right\}\right]\nonumber\\
  & +\hbar^{2}\frac{\tilde{g}}{8}\frac{\left(\tilde{t}+\tilde{g}\tilde{s}^{2}/2\right)^{2}}{(16\pi^2)^2}\left[\frac{4}{\epsilon^{2}}-\frac{4}{\epsilon}\left\{\gamma_{E} -1
     +\log{\left(\frac{\left(\tilde{t}+\tilde{g}\tilde{s}^2/2\right)}{4\pi}\right)}\right\}\right]-\hbar^{2}\frac{\tilde{g}^{2}\tilde{s}^{2}}{32\pi^{2}\epsilon}\left\{\frac{1}{4\pi}\theta^{(1)}\left(\frac{\left(\tilde{t}+\tilde{g}\tilde{s}^2/2\right)}{4\pi}\right)\right\}\nonumber\\
     & -\hbar^{2}\frac{\tilde{g}}{32\pi^{2}\epsilon}\left(\tilde{t}+\frac{\tilde{g}\tilde{s}^{2}}{2}\right)\left\{\frac{1}{4\pi}\theta^{(1)}\left(\frac{\left(\tilde{t}+\tilde{g}\tilde{s}^2/2\right)}{4\pi}\right)\right\},
\end{align}

where:

\begin{align}
        I_{1}[\tilde{m}^{2}] &=\frac{3}{(16\pi^2)^2}\int_{0}^{\infty}d\alpha\int_{0}^{\infty}d\beta \int_{0}^{\infty}d\gamma\frac{e^{-(\alpha+\beta+\gamma)\tilde{m}^2 }}{\mathcal{S}^{3}}\gamma(\beta-\gamma)\left[2\left(\vartheta^{4}\left(\frac{(\beta+\gamma)}{4\pi\mathcal{S}}\right)-1\right)+4\frac{(\beta+\gamma) }{4\pi\mathcal{S}}~\vartheta^{3}\left(\frac{(\beta+\gamma)}{4\pi\mathcal{S}}\right)\vartheta'\left(\frac{(\beta+\gamma)}{4\pi\mathcal{S}}\right)\right],\nonumber\\
        I_{2}[\tilde{m}^{2}] &=\frac{1}{(16\pi^2)^2}\int_{0}^{\infty}d\alpha\int_{0}^{\infty}d\beta\int_{0}^{\infty}d\gamma\frac{e^{-(\alpha+\beta+\gamma)\tilde{m}^2 }}{\mathcal{S}^{2}} \biggl[\left(\vartheta^{(2)}\left(0|\frac{i}{4\pi \mathcal{S}}\begin{bmatrix}
   \beta+\gamma & -\gamma\\
   -\gamma & \alpha+\gamma
   \end{bmatrix}\right)\right)^{4}-\vartheta^{4}\left(\frac{(\alpha+\beta)}{4\pi\mathcal{S}}\right)\nonumber\\
    &-\vartheta^{4}\left(\frac{(\beta+\gamma)}{4\pi\mathcal{S}}\right)-\vartheta^{4}\left(\frac{(\gamma+\alpha)}{4\pi\mathcal{S}}\right)+2\biggr],   
    \end{align}
    (with $\mathcal{S}=\alpha\beta+\beta\gamma+\gamma\alpha$)\\ 
    and:
    \begin{equation}
        A=-\frac{2}{\sqrt{3}}\int_{0}^{\frac{\pi}{3}}d\theta~\log{\left(2 \sin{\frac{\theta}{2}}\right)}.
    \end{equation}
Thus, the divergent term $\Gamma^{(div)}[\tilde{s}, L^{-1}]$ and the convergent term $\Gamma^{(conv)}[\tilde{s}, L^{-1}]$ halves of the quantity $\lim_{M\rightarrow\infty}\Gamma_{M}[\tilde{s}, L^{-1}]$ are respectively given by:

\begin{align}\label{div-rate2}
    \Gamma^{(div)}[\tilde{s}, L^{-1}] = \Gamma^{(div)}_{MF}+\Gamma^{(div)}_{1L}+\Gamma^{(div)}_{2L}-\log{[\mathcal{N}']},
\end{align}

and

\begin{align}\label{conv-rate2}
    \Gamma^{(conv)}[\tilde{s}, L^{-1}] = \Gamma^{(conv)}_{MF}+\Gamma^{(conv)}_{1L}+\Gamma^{(conv)}_{2L}.
\end{align}
 
    \twocolumngrid
\bibliography{main}
\bibliographystyle{apsrev4-1} 
\end{document}